\documentstyle[aps,preprint]{revtex}
 
\tighten

\newcommand{\be}{\begin{equation}} 
\newcommand{\ee}{\end{equation}}
\newcommand{\beq}{\begin{eqalignno}}
\newcommand{\eeq}{\end{eqalignno}} 
\newcommand{\epem}{\mbox{$e^+e^-$}}
 
\newcommand{\gev}{{\rm\,GeV}}
\newcommand{\tev}{{\rm\,TeV}} 
\newcommand{\fb}{{\rm\,fb}}

\newcommand{\mz}{M_Z}
\newcommand{\mw}{M_W} 
\newcommand{\cosb}{\cos\beta}
\newcommand{\sinb}{\sin\beta} 
\newcommand{\tanb}{\tan\beta}
\newcommand{\cosw}{\cos\theta_W} 
\newcommand{\sinw}{\sin\theta_W}

\newcommand{\chc}{\tilde{\chi}^{\pm}}
\newcommand{\chargino}{\tilde{\chi}^{\pm}_1}
\newcommand{\mchargino}{m_{\tilde{\chi}^{\pm}_1}}
\newcommand{\mch}{\mchargino} 
\newcommand{\chcp}{\tilde{\chi}^+}
\newcommand{\chcm}{\tilde{\chi}^-} 
\newcommand{\chn}{\tilde{\chi}^0}
\newcommand{\LSP}{\chn_1} 
\newcommand{\mLSP}{m_{\chn_1}}
 
\newcommand{\msnu}{m_{\tilde{\nu}}}
 
\newcommand{\mslep}{m_{\tilde{l}}}
 
\newcommand{\msq}{m_{\tilde{q}}}
\newcommand{\stot}{\sigma_{\rm total}}

\newcommand{\mtot}{{\cal M}^{\rm  tot}} 
\newcommand{\mprod}{{\cal M}^{\rm prod}}
\newcommand{\mdecay}{{\cal M}^{\rm decay}} 
\newcommand{\mum}{(\mu ,  M_2)}

\newcommand{\afbchi}{A_{FB}^{\chc}}
\newcommand{\afbjj}{A_{FB}^{jj}} 
\newcommand{\afbl}{A_{FB}^l}
\newcommand{\afbld}{A_{FB}^{l,\text{ decay}}}
 
\newcommand{\afbjjd}{A_{FB}^{jj,\text{ decay}}} 
\newcommand{\rhoplus}{\rho_{\chargino}}
\newcommand{\rhozero}{\rho_{\LSP}}

\newcommand{\ds}{\displaystyle}
 
\newcommand{\half}{\frac{1}{2}}

\newcommand{\tanw}{\tan\theta_W}

\newcommand{\rarr}{\rightarrow}
\newcommand{\sign}{{\rm sign}}

\begin{document}
\draft
\pagestyle{empty}

\preprint{
\begin{minipage}[t]{3in}
\begin{flushright}
LBNL--38954 \\
UCB--PTH--96/25 \\
RU--96--52 \\
hep-ph/9606477  
\end{flushright}
\end{minipage}
}
\title{Measuring SUSY Parameters at LEP II \\
Using Chargino Production and Decay
\vspace*{-0.2in}
}

\author{
Jonathan L. Feng
\thanks{Research Fellow, Miller Institute for Basic Research in
Science.}
\thanks{Work supported in part by the Director, Office of Energy
Research, Office of High Energy and Nuclear Physics, Division of High
Energy Physics of the U.S. Department of Energy under Contract
DE--AC03--76SF00098 and in part by the National Science Foundation under
grant PHY--90--21139.}
}
\address{
Theoretical Physics Group, LBNL and Department of Physics \\
University of California, Berkeley, California 94720
\vspace*{-0.1 in}}

\author{
Matthew J. Strassler
\thanks{Work supported in part by the Department of Energy under Contract
DE--FG05--90ER40559.}
}
\address{
Department of Physics and Astronomy\\
Rutgers University, Piscataway, New Jersey 08855
}
\maketitle
\vspace*{-0.2in}

\begin{abstract}\vspace*{-0.3in}

  Previously, in the context of the minimal supersymmetric standard
  model (without {\em a priori} assumptions of parameter unification),
  we studied the constraints on weak-scale SUSY parameters from chargino
  production at LEP II, using as observables $\mch$, $\mLSP$, the cross
  section and the leptonic branching fraction.  Here, exploiting the
  high degree of polarization in chargino production, we add to our
  earlier work the forward-backward asymmetries of the visible hadrons
  and leptons in chargino decays.  For a chargino that is mostly
  gaugino, the parameter space can now be restricted to a small region;
  $\tanb$ is constrained, the soft electroweak gaugino and electron
  sneutrino masses are determined to about 10\%, and the sign of $\mu$
  may be determined.  Constraints for a chargino that is mostly Higgsino
  are much weaker, but still disfavor the hypothesis that the chargino
  is mostly gaugino.  For a chargino which is a roughly equal mixture of
  Higgsino and gaugino, we find intermediate results.
\end{abstract}
 

\newpage
\pagestyle{plain}
\narrowtext

\setcounter{footnote}{0}

\section{Introduction}
\label{sec:intro}

One of the main goals of the LEP II $e^+ e^-$ collider at CERN
\cite{LEP200Overview} will be to search for signs of weak-scale
supersymmetry (SUSY) \cite{Reviews}.  If particles are discovered whose
quantum numbers suggest they might be SUSY partners of known particles,
the immediate issues will be to determine whether these particles really
behave in accordance with SUSY, and, if so, what values are assumed at
the electroweak scale by the parameters of the SUSY Lagrangian.  Among
the many new particles that might be found, the chargino, a mixture of
the $W$-fermion (Wino) and charged Higgs-fermions (Higgsinos), is of
particular interest.  From the theoretical standpoint, charginos are
expected to be lighter than gluinos, and in many models they are lighter
than most or all squarks and sleptons\cite{Peskin}.  If kinematically
accessible, charginos have a large cross section throughout SUSY
parameter space, produce a clean signal in certain decay modes, and have
properties that depend on a number of interesting SUSY parameters.  As
the chargino pair production cross section rises rapidly above
threshold, each step in collider energy holds the promise not only of
chargino discovery, but also of detailed SUSY studies from chargino
events.

Our goal in this paper, following on our earlier work \cite{PI}, is to
gain insight into the properties of charginos and to estimate the
ability of LEP II to determine the parameters of SUSY using
charginos.  This issue was also addressed by Leike in
Ref.~\cite{LeLEP}, where certain SUSY parameters were assumed to take
specific values, and in the work of Diaz and King\cite{DiazKing} in
the context of the five-parameter minimal supergravity scenario. Our
approach here is to avoid theoretical assumptions about physics at
high energy scales (such as the SUSY-breaking, GUT, or Planck scale),
and instead to exploit the fact that any given SUSY process is often
strongly sensitive to only a small subset of the SUSY parameters in
the electroweak-scale effective Lagrangian.  For chargino events, we
find that, after applying testable, phenomenologically-motivated
assumptions, only six parameters enter strongly.  Although the
resulting parameter space is still rather complicated, we have
developed effective methods in Ref.~\cite{PI} for understanding our
results and for displaying the qualitative relationship between
observables and the underlying SUSY parameters.  If charginos are
discovered, these methods may also prove very useful for interpreting
the correlated constraints on the parameter space that will come from
detailed global fits.

In our previous paper we considered the case where charginos decay
through virtual $W$-bosons, squarks and sleptons to the lightest
neutralino (which we assumed was stable or metastable) and two leptons
or light quarks.  Given this phenomenology, we investigated observables
that are independent of all angular distributions and thus can be
studied analytically at arbitrary beam energy.  These four observables
are the chargino mass $\mchargino$, the neutralino mass $\mLSP$, the
total cross section $\stot$ and the branching fraction $B_l$ of chargino
decays to leptons.  We showed that strong constraints on the weak-scale
parameters of SUSY often could be obtained from just these observables.
We considered, but did not use, other observables related to the angular
distribution of the production cross section.

In this paper, we use the fact that charginos are produced
predominantly by left-handed electrons \cite{LeLEP,Chia,LeNLC} to
obtain additional information.  The large polarization asymmetry
implies that, at the chargino production threshold, the two charginos
are usually produced with their spins aligned in the direction of the
positron momentum.  As we noted in Ref.~\cite{PI}, the angular
distributions of chargino decay products relative to the chargino spin
axis, which may be easily computed analytically, can then be measured,
and so may be used to obtain new information about the couplings at
the chargino decay vertices.  Of course, production rates very close
to threshold are too small for this to be useful, but since the
charginos are fermions, their cross section grows rapidly with energy
(Fig.~\ref{fig:csec}).  At the peak cross section, the energy is
sufficiently close to threshold that the angular distributions, though
modified, are still strongly correlated with the threshold angular
distributions, allowing our analytic techniques to be employed.

In Sec.~\ref{sec:assumptions} we review our main assumptions and the
parameter space under study.  Sec.~\ref{sec:chdecay} contains a
discussion of chargino decay amplitudes. In Sec.~\ref{sec:obsvs} we
discuss various observables and settle on the ones with the most new
information.  Our methods of generating events, extracting constraints
on parameter space from the observables, and representing these
constraints graphically are discussed in Sec.~\ref{sec:methods}.
Finally, we apply these ideas in Sec.~\ref{sec:studies} to the case
studies of our previous paper and present the resulting improvements
in the constraints on SUSY parameter space.

\section{Assumptions and the Six Parameters}
\label{sec:assumptions}

In Ref.~\cite{PI}, we made a number of assumptions about the
supersymmetric particle spectrum, leading to a specific phenomenology of
chargino production and decay controlled by six unknown SUSY parameters.
These assumptions are not terribly restrictive, in that they are obeyed
in large regions of the supersymmetric parameter space available to LEP
II.  Minor violations of these conditions generally lead to observable
effects in the data but do not completely invalidate our analysis, while
major violations would lead to qualitatively different observed
phenomena, requiring a separate study.  The motivations for these
assumptions, as well as possible violations thereof and methods for
detecting such violations, were fully discussed in Ref.~\cite{PI} and
will not be repeated here.  Instead, we will merely list the most
important assumptions and summarize their implications.

Our notation and conventions for the minimal supersymmetric standard
model are given in Appendix~\ref{app:MSSM}.  We assume the following.

\noindent (a) R-parity is conserved, so the lightest supersymmetric
particle is stable.  We assume this particle is either the lightest
neutralino, $\LSP$, or the gravitino.  In the latter case, we assume
the $\LSP$ is the next-to-lightest SUSY particle, and that it decays
to the gravitino outside the detector. If the $\LSP$ does
decay inside the detector to a photon and a gravitino, our analysis is
unaffected except for the nearly complete absence of standard model
background.

\noindent (b) The gaugino masses $M_1$ and $M_2$ and the $\mu$ 
parameter are independent quantities and are real, so there is no CP
violation in chargino processes.

\noindent (c) Sleptons, squarks and gluinos have masses beyond the 
kinematic limit of LEP II, and the intergenerational mixing in the
squark, slepton, and quark sectors is small enough to be neglected in
our analysis.  Decays through third-generation squarks and right-handed
sfermions are therefore suppressed either kinematically or by Yukawa
couplings. For the remaining sfermions, we assume the following
weak-scale mass relations (which define $\mslep$ and $\msq$):

\be\label{scalarassumption}
\begin{array}{rcl}
m_{\tilde{\nu}_{eL}}\approx m_{\tilde{e}_L} \approx &
m_{\tilde{\nu}_{\mu L}} & \approx m_{\tilde{\mu}_L} \approx
m_{\tilde{\nu}_{\tau L}}\approx m_{\tilde{\tau}_L}\approx  \mslep \\
m_{\tilde{u}_L}\approx & m_{\tilde{d}_L} & \approx
m_{\tilde{c}_L}\approx m_{\tilde{s}_L} \approx  \msq \ .
\end{array}
\ee
Note that approximate degeneracy of left-handed sfermions in the same
doublet is guaranteed, and intergenerational degeneracy is favored by
flavor-changing constraints.  As none of the other squarks and sleptons
play a role (under these assumptions) in chargino decay, only $\mslep$ and
$\msq$ appear as parameters in our analysis.

The main effect of these choices is to enforce a simple phenomenology,
under which charginos decay, via virtual $W$-bosons, squarks or
sleptons, to final states consisting of the lightest neutralino and
either two quarks or two leptons. The number of parameters controlling
the process is reduced to six: the basic SUSY parameters $\mu, M_1, M_2,
\tanb$ defined in Appendix~\ref{app:MSSM} and the slepton and squark
masses $\mslep, \msq$ defined above.  For reasons explained in Sec.~IIC
of Ref.~\cite{PI}, we take these parameters to lie in the ranges

\be\label{bounds}
\begin{array}{rcccl}
1       & \le  & \tanb & \le  & 50     \\
-1 \tev & \alt & \mu   & \alt & 1 \tev \\
0       & \le  & M_2   & \alt & 1 \tev \\
-M_2    & \le  & M_1   & \le  & M_2    \\
100 \gev& \le  &\mslep & \le  & 1 \tev \\
150 \gev& \le  &\msq   & \le  & 1 \tev \ .
\end{array}
\ee 

One of the most important issues regarding the lightest chargino is
whether it is largely gaugino or Higgsino.  As can be seen from
Eq.~(\ref{chamass}), this is determined by the three parameters $M_2$,
$\mu$ and $\tanb$, though the latter plays only a minor role.  If
$|\mu|\gg M_2$, then the light chargino is mostly gaugino; if $|\mu|\ll
M_2$, then it is mostly Higgsino.  We have chosen a measure of gaugino
content, namely $\rhoplus\equiv |{\bf V}_{11}|^2$, as a rough means of
dividing the $(M_2, \mu, \tanb)$ parameter space into a gaugino
($\rhoplus\geq.9$), Higgsino $(\rhoplus\leq.2)$ and mixed region, and we
present one case study from each region.  Limits will also be presented
on a measure of the gaugino content of the neutralino, $\rhozero \equiv
|{\bf N}_{11}|^2+|{\bf N}_{12}|^2$.

\section{Chargino Decay}
\label{sec:chdecay}

The amplitude for producing charginos that decay to a certain
final state,
\be\label{mtot}
\mtot = \sum _{s^+,s^- = -1} ^1 \mprod_{s^+s^-} \mdecay_{s^+}
\mdecay_{s^-} \frac{\pi}{\mchargino \Gamma_{\chargino}} \ ,
\ee
where

\be\label{mpieces}
\begin{array}{rl}
\mprod_{s^+s^-} &\equiv \mprod (\epem \rightarrow
\chcp_{s^+}\chcm_{s^-}) \ , \\
\mdecay_{s^{\pm}} &\equiv \mdecay(\chc_{s^{\pm}}
\rightarrow \LSP q'q, \LSP l\nu) \ ,
\end{array}
\ee 
and $s^{\pm}$ is (twice) the spin of $\chargino$ along the beam axis,
does not factorize into production and decay amplitudes.  However,
charginos are produced predominantly by left-handed electron initial
states \cite{LeLEP,Chia,LeNLC}, and the ratio $\sigma(e^-_R e^+_L \rarr
\chcp\chcm)/ \sigma(e^-_L e^+_R \rarr \chcp\chcm)$ does not exceed 15\%
in the accessible range of parameters\cite{PI}.  As an example, we plot
this ratio in Fig.~\ref{fig:sigmalr} for $m_{\tilde{\nu}} = 150 \gev$,
for which the ratio is nearly maximal. The production amplitude is
therefore dominated by the single spin component $\mprod_{s^+=s^-=1}$,
and the total amplitude in Eq.~(\ref{mtot}) approximately factorizes.
Furthermore, near threshold the production amplitude
$\mprod_{s^+=s^-=1}$ approaches a constant $\mprod_0$, independent of
production angle, plus corrections of order the chargino velocity $v$.
We may therefore write

\be\label{mtotapprox}
\begin{array}{rl}
  \mtot \approx \frac{\ds\pi\mprod_0}{\ds\mchargino
    \Gamma_{\chargino}} \left[\mdecay_{s^+=1}
  \mdecay_{s^-=1}\right]\Bigg|_{\ds s=4\mchargino^2} + {\cal O}(v) +
  {\cal O}(\mprod_{s^+=s^-=-1}) \ .
\end{array}
\ee
{}From this equation we see that properties of chargino
events near threshold, in particular all angular distributions,
are determined largely by the decay amplitude $\mdecay$.

As the beam energy is raised, however, the influence of $\mprod$ on
experimental observables increases, until ultimately, when the charginos
are highly relativistic, the angular distributions of their decay
products are completely determined by $\mprod$ and are insensitive to
$\mdecay$.  To maintain sensitivity to $\mdecay$, we must consider beam
energies close enough to threshold; however, to attain reasonable
statistics, we must run far enough above threshold.  The reader might be
concerned that these are mutually exclusive demands.  However, this is
not so, as we will see in Sec.~\ref{subsec:obsvundrafb}.  The
corrections that come from semi-relativistic chargino velocities are
indeed significant, and lead to loss of correlation between the measured
observables and their values at threshold. However, these corrections
are themselves correlated with other, already measured quantities, such
as the chargino mass, neutralino mass and chargino cross section.  Once
we measure these quantities and implement the constraints obtained in
Ref.~\cite{PI}, we find, using our Monte Carlo simulation, that the
correlations between the measured properties and the properties of the
decays tend to remain strong.  Thus, even though the measured quantity
will not equal the corresponding quantity formed from $\mdecay$, the
decay quantity may still be determined from the data and the Monte
Carlo, and our analytic methods can then be used.  This tends to confirm
that a global fit to data gathered well above threshold will still be
able to extract the information studied in this paper.

Under our assumptions, the amplitudes for a chargino to decay
to the lightest neutralino and either two leptons $\nu,e^+$ or two
quarks $u,\bar{d}$ can be written as \cite{decays,PI}
 
\be\label{decaytoall}
\frac{i^3g^2}{\sqrt{2}} \left[
\bar{u}(\chn_1)\gamma^\mu \big[D_L(F) P_L+D_R(F) P_R\big] u(\chcp_1)
\right]\left[\bar{u}(f)\gamma_\mu P_L v(\bar{f})\right] \ ,
\ee
where

\be\label{DLdefn}
D_L(F)=\frac{{\bf N}_{12} {\bf V}^*_{11}-\frac{1}{\sqrt{2}} {\bf
N}_{14} {\bf V}^*_{12} }{(p_f+p_{\bar{f}})^2-\mw^2}+
\frac{{\bf V}^*_{11}(Y_F \tanw {\bf N}_{11}+\half {\bf
N}_{12})} {(p_{\chn_1}+p_{f})^2-m_{\tilde F}^2} \ ,
\ee

\be\label{DRdefn}
D_R(F)=\frac{{\bf N}^*_{12} {\bf U}_{11}+\frac{1}{\sqrt{2}} {\bf
N}^*_{13} {\bf U}_{12}}{(p_f+p_{\bar{f}})^2-\mw^2}- \frac{{\bf
U}_{11}(Y_F \tanw {\bf N}^*_{11}-\half {\bf N}^*_{12})}
{(p_{\chn_1}+p_{\bar{f}})^2-m_{\tilde F}^2} \ .
\ee
Here ${\bf U}$, ${\bf V}$, and ${\bf N}$ are the chargino and
neutralino mixing matrices defined in Appendix~\ref{app:MSSM}, $Y_F$ is
the hypercharge of the left-handed fermion doublet $F$, and $m_{\tilde
F}$ is the mass of its superpartner.  For quarks, $F\equiv q$, $f\equiv
u$, $\bar{f}\equiv\bar{d}$, $Y_q=\frac{1}{6}$, while for leptons,
$F\equiv l$, $f\equiv \nu$, $\bar{f}\equiv e^+$, $Y_l=-\frac{1}{2}$.
Clearly the two terms in these expressions represent the virtual
$W$-boson and virtual sfermion diagrams.

For our analytic purposes, it is important to keep the momentum
dependence of the $W$-boson propagator in the amplitudes; however, the
effects of the squarks and sleptons can usually be well-approximated
by point propagators.  Corrections from this approximation have been
checked, using our Monte Carlo (in which the full propagators are
used), to be smaller than our experimental uncertainties, except for
the smallest values of the slepton masses.  For reference, the
energy-angle distributions for the charged lepton and for the dijet
system are presented in Appendix~\ref{app:EAdist}.  In most of the
allowed parameter space, $(\mch-\mLSP)^2\ll\mw^2$, so the
$D_{L,R}(l,q)$ are nearly constants independent of momenta.  We will
use this fact in our discussion below.

\section{Observables of Chargino Decays}
\label{sec:obsvs}

The next challenge that we face is to pick observables that can have
an impact on the determination of the weak-scale SUSY parameters.
Many observables of interest turn out to be correlated closely with
the already-determined quantities of Ref.~\cite{PI}, and therefore give no
additional information.  Fortunately, there are at least two new ones.

\subsection{Forward-Backward Asymmetries}
\label{subsec:AFBs}

The four complex functions $D_{L,R}(l,q)$ defined in
Sec.~\ref{sec:chdecay} determine the decay amplitudes.  In the
approximation that we ignore the momentum dependence of the propagators,
these quantities are constants.  After removing unobservable phases and
CP-violating phases (which we assumed were absent), four real quantities
remain.  We cannot determine the total width of the chargino (except in
extreme corners of parameter space) so the overall normalization of
these constants is unobtainable, but it is possible in principle to
extract their three independent ratios from the data.  The branching
fraction of charginos to leptons, $B_l$, which we have already used in
Ref.~\cite{PI}, gives us a parity-even combination of these ratios.  To
obtain the other two, $D_L(l)/D_R(l)$ and $D_L(q)/D_R(q)$, we need
parity-odd observables.  For leptons, we will use the forward-backward
asymmetry $\afbl$ in the angle $\theta_l$ between the charged lepton's
momentum and the direction of the positron beam.  For decays to quarks,
the cleanest variable is the forward-backward asymmetry $\afbjj$ in the
angle $\theta_{jj}$ between the momentum of the entire dijet system and
the positron axis; note that this quantity does not require a jet
definition and is insensitive to infrared effects, and so is relatively
free of systematic errors.

Exactly at threshold, and with perfect initial state polarization, so
that charginos are produced at rest in a definite spin state, these two
asymmetries are easily obtained by integrating the analytic formulas for
the differential decay rates of Appendix~\ref{app:EAdist} with respect
to the energy of the particle(s) in question. We will refer to these
threshold asymmetries as $\afbld$ and $\afbjjd$, to distinguish them
from the observed quantities $\afbl$ and $\afbjj$, which differ from the
former as a result of finite velocity, depolarization, cuts and detector
effects.  In the limit that $\mw$, $\msq$ and $\mslep$ are all much
greater than the chargino-neutralino mass difference, $\afbld$ and
$\afbjjd$ are functions only of the mass ratio $\mLSP/\mch$ and of the
relevant $D_L/D_R$.

To understand the type of information we can gain from these
measurements, it is useful to investigate the behavior of $D_L/D_R$ in
certain limits.  Let us first consider infinite squark and slepton
masses.  In this case $D_L(l)/D_R(l)=D_L(q)/D_R(q)$.  For $\tanb$ 
sufficiently large, $D_L/D_R \approx \mLSP/\mch$, and the
symmetry $\mu\leftrightarrow -\mu$ ensures that all relevant physical
quantities, under our assumptions, are functions only of $\mu^2$.
However, in the gaugino region, when

\be \tanb\ll \frac{\ds |\mu|}{\ds 2(M_2-M_1)} \ , \ee many matrix
elements depend linearly on $1/\mu$.  In particular this is true of
${\bf N}_{1i}$, $i=2,3,4$, and of ${\bf U}_{12}$ and ${\bf V}_{12}$.
It follows from Eqs.~(\ref{DLdefn}) and (\ref{DRdefn}) that, in this
region, $D_L/D_R$ will equal $\sign(M_1)+{\cal O}(M_W/\mu)$ and thus
will change substantially when the sign of $\mu$ is reversed.  In the
limit $\tanb=1$ the matrices ${\bf U}$ and ${\bf V}$ are equal and
${\bf N}_{i3}=-{\bf N}_{i4}$; furthermore, ${\bf N}_{i3}/{\bf N}_{i2}$
is real, so $D_L/D_R = {\bf N}_{12}/{\bf N}^*_{12} = \sign (M_1)$.
However this effect is in many cases invisible, as it applies only
very close to $\tanb=1$, is corrected outside the Higgsino region by
light squarks and sleptons, and may in the Higgsino region be
indistinguishable from the large $\tanb$ regime if
$\mLSP/\mch\approx 1$.

Finally, when sleptons [squarks] are sufficiently light, they will
begin to dominate $D_L(l)/D_R(l)$ [$D_L(q)/D_R(q)$] for large $|\mu|$.
In the limit $|\mu|\rarr\infty$ the $\LSP$ is pure hypercharge gaugino
while the $\chargino$ is pure Wino ($|{\bf U}_{11}|,|{\bf V}_{11}|,
|{\bf N}_{11}|\rightarrow 1$) so the $W$-boson diagram is completely
negligible, and therefore $D_L/D_R = -{\bf N}_{11}/{\bf N}^*_{11} =
-\sign (M_1)$; notice the all-important minus sign relative to the
infinite sfermion mass case.  The sfermion diagram dominates when

\be
|\mu|\gg
\frac{\sin 2\beta}{Y_F}\frac{\ds m_{\tilde F}^2}{M_2-M_1}
\ee
for small $\tanb$ and

\be
|\mu|\gg \frac{1}{\sqrt{2Y_F\tanw}}m_{\tilde F}
\ee
for large $\tanb$; here $Y_F$ is the hypercharge and
$m_{\tilde F}$ the mass of the slepton or squark.

We thus see several distinct regions in which the behavior of
$D_L/D_R$ can be characterized: the large $\tanb$ regime, the moderate
$|\mu|$ and low $\tanb$ regime where large subleading effects depend
on the sign of $\mu$, the regime $\tanb=1$, and finally the large
$|\mu|$ regime where the sleptons [squarks] dominate.  In other
regimes the behavior cannot be understood in simple analytic terms.
In Figs.~\ref{fig:afblG}, \ref{fig:afblH}, and \ref{fig:afbjjM}, where
asymmetries for various choices of fixed $\mchargino,$ $\mLSP$, and
either $\mslep$ or $\msq$ are plotted as a function\footnote{The
  utility of plotting quantities as a function of $(\alpha, \tanb)$
  was emphasized in Sec.~V A of Ref.~\cite{PI}.  This approach will be
  described more fully in Sec.~\ref{subsec:strategy} and
  Fig.~\ref{fig:sheets}.} of $\alpha=\tan^{-1}(M_2/\mu)$ and $\tanb$,
these regions easily can be identified: the first at the top center,
the second in the lower left and right corners, the third (only
visible in Fig.~\ref{fig:afblG}) at bottom center, and the last at the
far left and right.  (For example, in Fig.~\ref{fig:afblG}, the top
center has $\afbld< 5\%$, the lower left and right corners have
$\afbld\sim 10\%$ and $\afbld\sim 30\%$, the bottom center has
$\afbld\sim 20\%$ and the far left and right have $\afbld\sim35\%.$)
Since the variables considered in Ref.~\cite{PI} did not constrain
$\tanb$, the substantial $\tanb$ dependence of $D_L/D_R$ will provide
important new information.  The variation with the sign of $\mu$ will
allow us to rule out large regions in the gaugino case study below.
There is also dependence on the sign of $M_1$, which ensures that the
constraints for positive and negative $M_1$ will in general be quite
different, though it does not allow us to determine the sign of $M_1$.

It should be noted that two regimes with different $D_L/D_R$ might not
be distinguishable, because $\afbld$ and $\afbjjd$ are not monotonic
functions of $D_L/D_R$.  However, their functional dependence on
$D_L/D_R$ is not the same; for example, if
$|D_L(l)/D_R(l)|=|D_L(q)/D_R(q)|= 1$, $\afbjjd$ is
zero, but $\afbld$ depends on the sign of $D_L(l)/D_R(l)$. 

The existence of regions with qualitatively different sensitivity to
the underlying parameters, which is manifested in their widely varying
predictions for $\afbld$ and $\afbjjd$, is the first indication that
the inclusion of these observables will lead to significant
improvement over our previous results.

\subsection{Other Decay Observables}
\label{subsec:otherdobsv}

We will now discuss other possible observables, and explain why
we expect their impact on our analysis to be minor.

So far we have only used the angular distribution of the chargino
decay products, averaged over energy.  The energy distributions of
charged leptons and hadrons are obvious candidates for interesting
observables.  However, they in fact give very little additional
information.  The energy distributions are again sensitive, in the
limit $(\mch-\mLSP)^2\ll\mw^2, \msq^2, \mslep^2$, only to the
constants $D_L/D_R$ and $\mLSP/\mch$.  We have measured the first
through the asymmetries discussed above and have measured the second
directly.  Any additional information must stem from the variation of
$D_L/D_R$ with energy and angle.

In the case of the hadron energy-angle distribution (see Appendix
\ref{app:EAdist}) this variation can only give new information if the
squarks are very light. Furthermore, kinematics forces the hadron
energy range to be small (if $\mLSP\geq\half\mch$ the hadronic energy
varies by at most $\frac{1}{8}\mch$) and hadronic energy resolution
will therefore make any measurement highly imprecise.

In the case of the leptonic energy-angle distribution (see Appendix
\ref{app:EAdist}) the variation of $D_L(l)/D_R(l)$ can tell us
something about $\mslep$, unless we are in the Higgsino region, where
the sleptons essentially decouple.  While this might help distinguish
the mixed or gaugino region from the Higgsino region, it will not do
much more; a good measurement of the slepton mass is already achieved in
these regions using only $\stot$ and $B_l$ \cite{PI}.  Furthermore, the
leptonic energy distribution tends to be highly peaked near its
midpoint, and small chargino-neutralino mass splitting, which would be
present in the Higgsino region, would reduce any effect by limiting the
range of lepton energies.  We therefore expect relatively little
additional information from this observable.

One might also consider energy-angle distributions of individual quark
jets.  This has many systematic problems (having to do with jet
definitions and energy resolution) and again is a function only of
$D_L(q)/D_R(q)$ and $\mLSP/\mch$ unless the squarks are very light.  We
therefore do not think observables based on these distributions will add
very substantially to our analysis.

However, one should not conclude that these observables
are completely uninteresting, only that they do not impact our
present work, in which we have made certain assumptions
about the phenomenology.  In fact, they may be used
to test our assumptions.  Should the decay amplitudes be more
complicated than Eqs.~(\ref{decaytoall})--(\ref{DRdefn}) so that
other Lorentz structures are introduced (by the presence, for
example, of charged Higgs bosons or right-handed squarks in the
decays) then new parameters will enter, and their effects
will be distinguishable from those of merely changing the values of 
$D_{L,R}(l,q)$.

\subsection{Production Observables: $\stot(s)$, $\afbchi$, $A_{LR}$}
\label{subsec:otherpobsv}

Although we will not use them here, three other observables are also
of interest.  The first of these is the energy dependence of the cross
section which can be obtained by a straightforward beam scan.  That
this is an interesting quantity is hinted at by Fig.~\ref{fig:csec},
where it can be seen that the shape of the curves depends
on the underlying parameters.  The second is the forward-backward
asymmetry $\afbchi$ of chargino production.  Both of these are
sensitive to the presence of a light electron sneutrino and might
serve to distinguish the Higgsino region from the mixed region, which
can otherwise look quite similar.

At threshold $\afbchi$ is zero, so all asymmetries of chargino decay
products are due to parity violation in the decay amplitudes.  By
contrast, as noted earlier, at ultra-relativistic energies all
chargino decay products travel in the direction of the chargino, so
all observed asymmetries are due to the production amplitude and
$\afbchi$ is easily measured. In short, the observed $\afbjj$
corresponds near threshold to the asymmetry $\afbjjd$
and at high energy to the asymmetry $\afbchi$.  
Chargino velocities at LEP II will be at most
semi-relativistic, so no simple measurement of $\afbchi$ can be made.
In fact, at these energies, as argued in Ref.~\cite{PI} and
demonstrated more clearly below, most of the information available in
chargino angle asymmetries comes from the decay amplitude.  However,
by varying the beam energy, one can in principle separate the
contribution to the asymmetries from decay and production amplitudes.

Thus, an energy scan will permit both a measurement of 
$\stot(s)$ and a clearer separation of $\afbjjd$ and $\afbchi$.
We have not studied the question of determining the optimal approach 
for such a scan or whether it would substantially improve
the determination of underlying parameters; should charginos be found 
this issue will require investigation.

A third interesting observable is the left-right production
asymmetry 
$A_{LR} = 
\sigma(e^-_Le^+_R\rarr \chcp\chcm)/
\sigma(e^-_Re^+_L\rarr\chcp\chcm)$, 
but unfortunately we do not know of an efficient way to measure this at
LEP II.  There is sensitivity to this variable in the correlations
between lepton and hadron angles, but since the polarization is expected
always to be at least 85\%, we do not expect much to come of this
variable.  Of course, as a test of SUSY, it should be checked that
hadron-lepton correlations are consistent with the expected
polarization.

\section{Methods}
\label{sec:methods}

\subsection{Simulations}
\label{subsec:expmethods}

Since the two chargino decays are independent, chargino pair
production leads to a final state of two neutralinos plus four
leptons, two leptons and two quarks, or four quarks.  We refer to
these three modes as leptonic, mixed and hadronic.  For measurements
of the asymmetries $\afbjj$ and $\afbl$, the mixed mode events (which
appear in the detector as one charged lepton, two or more jets, and
missing energy) are by far the best.  These events have a clean
signature, low backgrounds\cite{Aachen,Chen,Grvz2}, and a well-studied
and understood set of cuts\cite{Grvz2}.  By contrast, hadronic mode
events are plagued by the difficulty of determining which jets come
from the $\chcp$ and which from the $\chcm$, while leptonic mode
events, where only two acoplanar charged leptons are visible, may have
substantial backgrounds from $W^+W^-$ production.

For this study, chargino events were generated using a simple parton
level Monte Carlo event generator with all spin correlations included.
Hadronization and detector effects were crudely simulated by smearing
the parton energies with detector resolutions currently available at
LEP: $\sigma_E^{\text{ had}}/E = 80\%/\sqrt{E}$ and $\sigma_E^{\text{
e.m.}}/E = 19\%/\sqrt{E}$, with $E$ in GeV.  Initial state radiation was
not included.  To extract the mixed mode chargino signal from the
standard model background, the cuts of Ref.~\cite{Grvz2} were employed.
Additional details concerning the event simulation may be found in
Ref.~\cite{PI}, where the final uncertainties in observables were shown
to be fairly insensitive to the experimental assumptions.

If, as has been suggested in Refs.~\cite{Fermievent,DSBsignals}, each
neutralino decays in the detector to a photon and a gravitino, the two
outgoing photons simply serve to tag the otherwise-unchanged chargino
event.  One may therefore simply dispense with most cuts since there
will be no important standard model background to chargino events.  The
rest of our analysis is unaffected; the information gained from chargino
events is roughly the same, except that lower integrated luminosity is
required.

\subsection{Observed Asymmetries and Underlying Asymmetries}
\label{subsec:obsvundrafb}

Although our analytic work is appropriate for the threshold region, it
is clear that the best results are to be found somewhat above threshold.
While our formulas become less accurate as the chargino velocity
increases, the increasing cross section gives considerably better
statistics.  Our previous study was done at a center-of-mass energy of
190 GeV.  It turns out that for a chargino of mass 80 GeV (as in the
case studies below) this generates nearly the maximum rate, as shown in
Fig.~\ref{fig:csec}, and for simplicity (since much of our previous work
can be carried over) we will do our analysis there.  We have found that
reducing the energy to 170 GeV does not greatly improve our results.
Our formulas, although numerically inaccurate as the beam energy is
raised, are still well-correlated with the observed asymmetries when the
other observables ($\mchargino, \mLSP, \stot, B_l$) are held fixed.

In each of our case studies, we study the constraints stemming from an
integrated luminosity of $1 {\text{ fb}}^{-1}$.  Using our Monte Carlo
program as described above, we apply realistic cuts to extract the
chargino signal in the mixed mode.\footnote{Strictly speaking we are
  using the ``Y mode'', as discussed in detail in Ref.~\cite{PI}; this
  means that we only use those tau lepton events in which the tau
  decays leptonically.} 
These events show 
forward-backward asymmetries $\afbl$ and $\afbjj$ in the lepton and
hadronic angles.  From these raw asymmetries we must determine the
underlying asymmetries $\afbld$ and $\afbjjd$ relative to the chargino
spin axis, which are those which would be measured at threshold given 
100\% polarization of the chargino spins.

To do this, we run a number of simulations with parameters that give
the same values for the observables ($\mch,\mLSP,\stot,B_l$) but which
have different values of $\afbld$ ($\afbjjd$).  The observed $\afbl$
($\afbjj$) differs from $\afbld$ ($\afbjjd$), but is correlated with it.
The correlation is approximately linear (as will be seen in a particular
case below) which makes it possible to measure the underlying asymmetry
using the observed one.  However, the correlation suffers from some
smearing due to imperfect polarization of the charginos and due to their
non-zero velocities combined with the production asymmetries.  There is
also smearing due to angle-dependent variations in efficiency which are
associated with our cuts and with detector resolution.  This smearing
introduces some additional uncertainty in the determination of $\afbld$
and $\afbjjd$, which we will treat as systematic error, and which we
account for in our error estimates.  In each of our studies, the
systematic effects are smaller than or of the same order as statistical
uncertainties.

As an example, consider the gaugino case study discussed in
Sec.~\ref{subsec:gaugino}.  The observed hadronic asymmetry $\afbjj$ is
determined up to experimental statistical error.  To determine the
underlying $\afbjjd$, we must first determine the degree of correlation
between $\afbjj$ and $\afbjjd$.  We choose a large number of points in
SUSY parameter space with values of $\mchargino$, $\mLSP$, $\stot$ and
$B_l$ that are within experimental errors of those of the underlying
gaugino point.  For each of these, we run a Monte Carlo simulation, and
compare the values of $\afbjj$ and $\afbjjd$; the results (which are
exceptionally good in this case study) are plotted in
Fig.~\ref{fig:afbcor}, where each point represents a set of SUSY
parameters.  The deviation from perfect correlation is caused by two
effects: Monte Carlo statistical error and the systematic error
discussed above.  The Monte Carlo statistical error is removed to
determine the systematic error, and finally the total error expected for
$\afbjjd$ is determined by adding to the systematic error the
experimental statistical error. For simplicity, we combine the
statistical and systematic errors in quadrature.  Readers interested in
the details of the error analysis are referred to the Appendix of
Ref.~\cite{whetherSUSY}.

The high degree of correlation in the case of Fig.~\ref{fig:afbcor} is
due in part to the fact that the production asymmetry $\afbchi$ is
already determined by the other observables to be 12--21\% (see Fig.~20
of Ref.~\cite{PI}).  In the Higgsino and mixed case studies, $\afbchi$
lies between $0$ and $20\%$, so the correlation between $\afbjj$ and
$\afbjjd$ is somewhat less impressive and the systematic uncertainties
are larger.  Still, $\afbjjd$ varies over a much wider range and so is
responsible for most of the potential variation in the observed
quantity; even with the large systematic errors it puts strong
restrictions on the parameter space.  In fact, the systematic error we
obtain is misleadingly large.  The part of it which is due to cuts and
detector effects cannot be removed, but a certain fraction of it stems
from our use of formulas appropriate for the threshold region in a
semi-relativistic regime.  A full global fit to the data will of course
use the exact matrix element and kinematics, as well as additional
kinematic information not used by us, and may thereby reduce the
systematic error substantially.

\subsection{Strategy for Finding Allowed Parameter Space}
\label{subsec:strategy}

After the quantities $\mchargino$, $\mLSP$, $\stot$, $B_l$, $\afbjjd$
and $\afbld$ are measured, one must determine how the six-dimensional
SUSY parameter space is restricted.  To characterize the favored regions
of parameter space, we define ${\cal R}_1$ (${\cal R}_2$) to be the
subspace of parameter space in which all predicted observables lie
within one (two) standard deviations of the actual observations.  These
regions contain the points in parameter space most consistent with the
measurements, and we will refer to ${\cal R}_1$ and ${\cal R}_2$ as the
inner and outer allowed regions, respectively.

These allowed regions are complicated six-dimensional spaces, and of
course we are only able to display their projections onto
lower-dimensional subspaces.  In this study, following the methods of
Ref.~\cite{PI}, we choose to display our results in two ways.  First, we
consider one-dimensional projections: for each parameter, we give its
``global bounds'', which we define to be its range within ${\cal R}_1$.
In addition, we will also present the two-dimensional projections of
${\cal R}_1$ and ${\cal R}_2$ onto the plane ${\cal T}$ \cite{PI}, which
we now describe.

In defining the plane ${\cal T}$, the dependence of $\mchargino$ on only
three SUSY parameters allows us a simple starting point.  First,
consider the three-dimensional space $(\mu, M_2, \tanb)$.  When the
chargino mass is measured, it constrains the allowed region to lie in
two thin sheets, which we will label as $\cal S$, one with $\mu<0$ and
another with $\mu>0$. This is shown schematically in
Fig.~\ref{fig:sheets}.  The two sheets $\cal S$ may then be flattened
into a plane $\cal T$ with the coordinate transformation

\be 
(\mu, M_2, \tanb ) \in {\cal S} \rightarrow \left(\alpha\equiv
\arctan \frac{M_2}{\mu}, \tanb \right) \in {\cal T} \ , 
\ee 
as shown in Fig.~\ref{fig:sheets}. Since the sheets are not infinitely
thin, a short segment of points in $\cal S$ is projected into every
point in $\cal T$.  We will also refer to $\cal T$ as the $(\alpha,
\tanb)$ plane.  Note that by presenting results on ${\cal T}$,
constraints on $\tanb$ and the gaugino content of the chargino are
easily understood.  The far gaugino regions are transformed to the
areas with $\alpha \approx 0^\circ , 180^\circ$, and the far Higgsino
regions now correspond to the region with $\alpha \approx 90^\circ$.
Note also that the symmetry $\mu\leftrightarrow -\mu$ for
$\tanb\rarr\infty$ implies that, at large $\tanb$, observables at
$\alpha$ are nearly equal to those at $180^\circ-\alpha$.

The global bounds and allowed regions in the $(\alpha, \tanb)$ plane
cannot be associated with definite confidence levels, and are only
meant to give a rough idea of the constraints on parameter space from
the various observables.  Ideally, we would determine the probability
distribution in parameter space, display various projections of this
probability distribution, and determine the regions bounded by several
different values of $\chi^2$.  Such a procedure requires detailed
knowledge of the correlations between the various measurements,
however, and is beyond the scope of this analysis.  Here, we simply
note that, were all six measurements uncorrelated, the probability
that all of them would lie within $1\sigma$ ($2\sigma$), that is, the
probability that the underlying physical parameters would lie within
region ${\cal R}_1$ (${\cal R}_2$), is 10\% (73\%).  Of course, the
probability that any single global bound holds, or that the parameters
$(\alpha, \tanb)$ lie in an allowed region irrespective of other
parameters, is much larger than the probability of lying in the
associated region ${\cal R}_1$ or ${\cal R}_2$. For example, if a
given parameter is primarily constrained by only one observable, that
parameter's global bound is roughly a 1$\sigma$ (68\% C.L.) bound.

Although we have argued that the shape of the allowed regions within the
$(\alpha,\tanb)$ plane gives considerable information, we are only
looking at two dimensional projections, and some of the structure is
necessarily lost.  Additional insight into the structure of these
regions may be obtained by using the ``minmax'' plots described in our
earlier paper \cite{PI}.  For example, the neutralino mass $\mLSP$ is a
function of $\mu$, $M_2$, $\tanb$, and $M_1$, and so the $\mLSP$
measurement limits $M_1$ to a certain range for each point in $\cal T$.
In Ref.~\cite{PI}, to represent graphically this restriction of $M_1$,
or equivalently, $M_1/M_2$, we did the following.  For a point ${\cal P}
= (\alpha,\tanb) \in {\cal T}$, we found all parameters $(\mu,
M_2=\mu\tan\alpha, \tanb, M_1/M_2)$ such that the corresponding values
of $\mchargino$ and $\mLSP$ were within one standard deviation of their
observed values.  The allowed values of $M_1/M_2$ lay in some range
$(M_1/M_2)_{\rm min} < M_1/M_2 < (M_1/M_2)_{\rm max}$.  To display this
range, we plotted contours in $\cal T$ of $(M_1/M_2)_{\rm min}$ and
$(M_1/M_2)_{\rm max}$. (We refer to this as a ``minmax plot''.)  In a
similar manner, the measurement of $\stot ( \mu, M_2,
\tanb, \mslep)$ limited the allowed range of $\mslep$ and the measured
value of $B_l( \mu, M_2, M_1/M_2,\tanb, \mslep,\msq)$ restricted the
range of $\msq$.  Improved restrictions on these parameters as a
function of $(\alpha, \tan\beta)$ may be crudely estimated by
overlaying the new, smaller allowed regions presented here onto the
minmax plots of Ref.~\cite{PI}, though the actual restrictions will be
somewhat tighter.  In the gaugino case study of the next section, we
will present a new minmax plot for $\msq$.

\section{Case Studies}
\label{sec:studies}

Having discussed the relevant observables in chargino pair production,
we now consider their effectiveness in constraining SUSY parameter space
in three specific examples.  We consider one case study in each of the
three regions of parameter space.

\subsection{Gaugino Region}
\label{subsec:gaugino}

We turn first to our case study in the gaugino region.  The
underlying parameters are taken to be

\be\label{gparam}
(\mu, M_2, \tanb, M_1/M_2, \mslep, \msq) = (-400,75,4,0.5,200,300) \ ,
\ee
giving $\alpha = 169^\circ$.  Given an integrated luminosity of 
$1\text{ fb}^{-1}$, there are 3200 chargino events, of which 1246 are 
mixed mode (technically, ``Y mode'' \cite{PI}), and of these, 968 (78\%) 
pass the cuts.  As shown in Ref.~\cite{PI}, the four original 
observables can be determined with uncertainties

\be\label{gvalues}
\begin{array}{rcl}
\mchargino &=& 79.6 \pm 2.7 \gev \\
\mLSP      &=& 39.1  \pm 2.3 \gev \\
\stot      &=& 1.16 \pm .06\ {\rm R} = 3200\pm 160 \fb \\
B_l        &=& 0.42 \pm .02 \ , \\
\end{array}
\ee
where both systematic and statistical errors have been included. These
constraints lead to impressive restrictions on the parameter space:
$\stot$ and $B_l$ force $\alpha$ to lie in the gaugino region, while
$\mLSP$ and $\stot$ also constrain $M_1/M_2$ and $\mslep$.  However,
neither $\tanb$ nor $\msq$ are well constrained.  Below, we will find
significant improvements in the determination of these parameters once
the decay asymmetries are considered.

The measured values of $\afbl$ and $\afbjj$ are determined up to
statistical errors to be $8.0\pm3.2\%$ and
$22.8\pm3.1\%$.  Applying the method described above we determine
the underlying asymmetries to be

\be\label{GAg}
\afbld = -0.6\pm4.8 \% \ ; \ \afbjjd = 16.9\pm5.1 \% \ ,
\ee
where both statistical and systematic errors are included.  In this case
study the uncertainties are dominated by statistical errors. For
simplicity we take the central values to be the actual values.

To see the impact of the $\afbld$ measurement, consider
Fig.~\ref{fig:afblG} which shows $\afbld$ as a function of $\alpha$ and
$\tanb$ for $M_1>0$ and $\mch$, $\mLSP$, and $\mslep$ fixed to the
values appropriate to this case study.  At one standard deviation, the
$\afbld$ measurement prefers a small region at $\tanb < 7$ and
$\alpha>160^{\circ}$; our previous results had no $\tanb$ restrictions
and favored both $\alpha<20^{\circ}$ or $\alpha>160^{\circ}$.  In short,
only the gaugino region for large and negative $\mu$ and for small or
moderate $\tanb$ is consistent with this value of $\afbld$.  (For
negative $M_1$ there is a similar result but the allowed region extends
to somewhat larger values of $\tanb$.)

We first present global bounds on each parameter separately, as
described in Sec.~\ref{subsec:strategy}.  These are
\be\label{grangespos}
\begin{array}{rcccl}
1.0       &<& \tanb           &<& 6.1 \\
-1\tev    &<&  \mu     &<&  -290 \gev \\
68\gev  &<& M_2             &<& 80\gev \\
34\gev  &<& M_1             &<& 40\gev \\
0.44  &<&  \frac{M_1}{M_2}  &<& 0.57 \\
0.99    &<& \rhoplus        &<& 1.00 \\
0.98    &<& \rhozero        &<& 1.00 \\
180\gev &<& \mslep          &<& 225 \gev \\
150\gev &<& \msq            &<& 1\tev \ 
\end{array}
\ee
for positive $M_1$ and
\be\label{grangesneg}
\begin{array}{rcccl}
1.0      &<& \tanb           &<& 24.2 \\
-1\tev    &<&  \mu     &<&  -295 \gev \\
70\gev  &<& M_2             &<& 85\gev \\
-44\gev  &<& M_1             &<& -37\gev \\
-0.62  &<&  \frac{M_1}{M_2} &<&  -0.45 \\
0.99    &<& \rhoplus        &<& 1.00 \\
0.98    &<& \rhozero        &<& 1.00 \\
180\gev &<& \mslep          &<& 225 \gev \\
150\gev &<& \msq            &<& 1\tev \ 
\end{array}
\ee 
for negative $M_1$.  As shown in Ref.~\cite{PI}, the results strongly
favor gaugino mass unification, as well as the hypothesis that the
chargino is nearly pure gaugino.  In addition, however, the
determination of $M_2$ and $M_1$, the fixing of sign$(\mu)$, and the
limits on $\tanb$ are quite strong, and represent significant
improvements over the constraints found in Ref.~\cite{PI}.

The allowed regions for positive and negative $M_1$ are shown in
Figs.~\ref{fig:Gallowedp} and \ref{fig:Gallowedm}.  Even the outer
contour lies in a very small range of $\alpha$, and, for positive $M_1$,
of $\tanb$ as well.  For negative $M_1$ the outer contour ends at
$\tanb\sim 38$.

Although the squark masses are not constrained globally, they are
still strongly correlated with $\alpha$ and $\tanb$.  To illustrate
this, we present minmax plots for $\msq$ in Figs.~\ref{fig:msqminmaxp}
and \ref{fig:msqminmaxm}, which show this correlation inside the
inner contour.  These should be read as follows: at any point
$(\alpha,\tanb)$ in the inner allowed region, the upper (lower) plot
gives the minimum (maximum) value of $\msq$ in ${\cal R}_1$.  (Recall
the inner allowed region is the projection of ${\cal R}_1$ onto the
$(\alpha,\tanb)$ plane.)  It is evident that low $\tanb$ and larger
values of $\alpha$ prefer low $\msq$.

\subsection{Higgsino Region}
\label{subsec:Higgsino}

The next case study, in the Higgsino region, has as underlying
parameters

\be\label{hparam}
(\mu, M_2, \tanb, M_1/M_2, \mslep, \msq) = (-75,250,4,0.5,200,300) \ ,
\ee
for which $\alpha = 107^\circ$.  Of 2450 chargino events, 892 are mixed
mode, and 269 (30\%) of these pass the cuts.  The observables and their
uncertainties as determined in Ref.~\cite{PI},

\be\label{hvalues}
\begin{array}{rcl}
\mchargino &=& 79.7 \pm3.0 \gev \\
\mLSP      &=& 62.3 \pm2.6 \gev \\
\stot      &=& 0.89 \pm 0.10\ {\rm R} = 2450 \pm 250\fb \\
B_l        &=& 0.34 \pm .05 \ , \\
\end{array}
\ee
do not lead to very strong restrictions on the parameters.  While
$\mLSP$ leads to significant correlations between $\alpha$ and $M_1$,
and while $\stot$ restricts $\mslep$ to be low in the allowed part of
the gaugino and mixed region, the constraints on the $(\alpha, \tanb)$
plane are not very impressive.  Essentially (see Fig.~25 in
Ref.~\cite{PI}) the allowed region lies between
$10^\circ<\alpha<170^\circ$, with almost no correlation with $\tanb$.
We will see some slight improvement in this situation below.

The measured values of $\afbl$ and $\afbjj$ and their statistical errors
are $-11.2\pm6.1\%$ and $-10.7\pm6.1\%$, from which we extract the
underlying values

\be\label{GAh} 
\afbld = -4.3\pm9.5 \% \ ; \ \afbjjd = -13.3\pm11.4 \%
\ .  
\ee 
In this case study the correlation between the observed and
underlying values of the asymmetries has much larger systematic error
than the previous one (largely due to the greater variation of the
underlying $\afbchi$) and the statistical and systematic errors
contribute almost equally.  The errors could therefore be reduced
somewhat by a global fit or by working closer to threshold; we have
checked however that reducing the beam energy to $170$ GeV does not
change our results substantially.

Fig.~\ref{fig:afblH} shows $\afbld$ for $M_1>0$ and
fixed $\mchargino=80\gev$,
$\mLSP = 62\gev$, and $\mslep=175\gev$ plotted as a function of
$\alpha$ and $\tanb$.  (We choose  $\mslep=175\gev$ because, although 
the Higgsino region is completely insensitive to $\mslep$, the allowed
portion of the gaugino region already requires this value in order to match
the observed cross section.)  The $\afbld$ measurement prefers
$25^\circ<\alpha<155^\circ$, except for a part of the low
$\tanb$ gaugino region at large $\alpha$.  The $\afbjjd$ measurement
pushes this latter region, as well as the $\tanb\sim 1$ part of the
mixed and Higgsino region, outside the inner allowed contour.

The global bounds (see Sec.~\ref{subsec:strategy}) on the parameters are
\be\label{hrangespos}
\begin{array}{rcccl}
1.2       &<& \tanb           &<& 50 \\
-220\gev   <  \mu              < -60 \gev
&\quad& {\rm or} &\quad&
80 \gev <  \mu              <  230 \gev \\
75\gev  &<& M_2             &<& 1\tev \\
60\gev  &<& M_1             &<& 360\gev \\
0.12 &<&  \frac{M_1}{M_2} &<& 1.00 \\
0.01    &<& \rhoplus        &<& 0.98 \\
0.00    &<& \rhozero        &<& 0.93 \\
100\gev &<& \mslep          &<& 1\tev \\
150\gev &<& \msq            &<& 1\tev \ 
\end{array}
\ee
for positive $M_1$ and
\be\label{hrangesneg}
\begin{array}{rcccl}
1.4       &<& \tanb           &<& 50 \\
-130\gev   <  \mu              < -65 \gev
&\quad& {\rm or} &\quad&
80 \gev <  \mu              <  135 \gev \\
95\gev  &<& M_2             &<& 1\tev \\
-365\gev&<& M_1             &<& -65\gev \\
-1.00 &<&  \frac{M_1}{M_2} &<& -0.08 \\
0.01    &<& \rhoplus        &<& 0.72 \\
0.00    &<& \rhozero        &<& 0.87 \\
100\gev &<& \mslep          &<& 1\tev \\
150\gev &<& \msq            &<& 1\tev \ 
\end{array}
\ee
for negative $M_1$.  The most impressive of these are $|\mu|<230$ GeV,
which  disfavors the far gaugino region, and
$60\gev<|M_1|<365\gev$.  Also interesting is that $\tanb\sim 1$ is
disfavored; recall that $D_L/D_R$ changes quickly from $1$ at $\tanb=1$
to $\mLSP/\mchargino$ at slightly larger values of $\tanb$, so that with
sufficient statistics these two subregions may be distinguished
experimentally.

In Figs.~\ref{fig:Hallowedp} and \ref{fig:Hallowedm}, the allowed region
is shown for positive and negative $M_1$.  (Its complicated structure is
due to combining constraints from $\mLSP$ and $\afbjjd$.)  The inner
contour runs between $25^\circ<\alpha<155^\circ$; parts of the far
gaugino region and $\tanb=1$ still lie inside the outer contour but are
not favored.

The absence of global bounds on $\mslep$ is to be expected, since in
the Higgsino region all observables are independent of this quantity.
However, there are important bounds on $M_1/M_2$ and $\mslep$ for
$|\mu|>M_2$, which may be estimated by overlaying the allowed region
on the minmax plots Figs.~22 and 23 of Ref.~\cite{PI}. This method gives an
underestimate of the constraints since only $\mLSP$ and $\stot$ are
used in those plots, but it can be seen that gaugino mass unification
($M_1/M_2\approx .5$) is disfavored in the gaugino and far Higgsino
region due to the ratio $\mLSP/\mch$, and that the allowed parts of 
the mixed and gaugino region require a light slepton due to the
low $\stot$.

\subsection{Mixed Region}
\label{subsec:mixed}

Our final case study lies in the mixed region, with underlying
parameters  

\be\label{mparam}
(\mu, M_2, \tanb, M_1/M_2, \mslep, \msq) = (-90,115,4,0.5,200,300) \ ,
\ee
for which $\alpha = 128^\circ$.  Of 2070 chargino events, 718 are mixed
mode, of which 428 (60\%) pass the cuts.  As discussed in
Ref.~\cite{PI}, the measurements
\be\label{mvalues}
\begin{array}{rcl}
\mchargino &=& 80.3 \pm 3.3\gev \\
\mLSP      &=& 52.8 \pm2.7 \gev \\
\stot      &=& 0.75 \pm 0.05\ {\rm R} = 2070 \pm 130\fb \\
B_l        &=& 0.32 \pm 0.03\\
\end{array}
\ee
disfavor both the far Higgsino and far gaugino regions, though with
relatively little $\tanb$ dependence.  A measurement of the slepton mass
was achieved, while no other parameters were well-determined.

The experimentally measurable asymmetries and their statistical
uncertainties for $\afbl$ and $\afbjj$ are $-8.8\pm4.8\%$ and
$-20.5\pm4.7\%$, from which we extract the underlying values

\be\label{GAm}
\afbld = -5.9\pm7.7 \% \ ; \ \afbjjd = -28.6\pm10.0 \% \ .
\ee
As in the previous case study, the statistical and systematic errors are
nearly equal, due to large systematic uncertainties in the correlation
between the observed and measured asymmetries.

Fig.~\ref{fig:afbjjM} shows $\afbjjd$ for $M_1>0$ and fixed
$\mchargino=80\gev$, $\mLSP = 53\gev$, and $\msq=300\gev $ as a
function of $\alpha$ and $\tanb$.   This observable has interesting
dependence on $\tanb$ and on sign$(\mu)$; it rules out the far gaugino
region as well as disfavoring the $\alpha<90^\circ$ low $\tanb$
region.

The global bounds (see Sec.\ref{subsec:strategy})
on the underlying parameters are

\be\label{mrangespos}
\begin{array}{rcccl}
1.0       &<& \tanb           &<& 50 \\
-175\gev   <  \mu              < -55 \gev
&\quad& {\rm or} &\quad&
85 \gev <  \mu              <  155 \gev \\
55\gev  &<& M_2             &<& 560 \gev \\
40\gev  &<& M_1            &<& 90 \gev \\
0.16 &<&  \frac{M_1}{M_2} &<& 0.95 \\
0.05    &<& \rhoplus        &<& 0.94 \\
0.06    &<& \rhozero        &<& 0.90 \\
100\gev &<& \mslep          &<& 260 \gev \\
150\gev &<& \msq            &<& 1\tev \ 
\end{array}
\ee
for $M_1>0$ and 

\be\label{mrangesneg}
\begin{array}{rcccl}
2.5      &<& \tanb           &<& 50 \\
-110\gev   <  \mu              < -85 \gev
&\quad& {\rm or} &\quad&
85 \gev <  \mu              <  135 \gev \\
140\gev  &<& M_2             &<& 560 \gev \\
-75\gev  &<& M_1            &<& -50 \gev \\
-0.44 &<&  \frac{M_1}{M_2} &<& -0.12 \\
0.05    &<& \rhoplus        &<& 0.65 \\
0.54    &<& \rhozero        &<& 0.89 \\
100\gev &<& \mslep          &<& 230\gev \\
150\gev &<& \msq            &<& 1\tev \ 
\end{array}
\ee for $M_1<0$.  Most impressive are $|\mu|<155$ GeV,
$40<|M_1|<90\gev$, $M_2<530$ GeV, and $\mslep<225$ GeV, though only
the first two represent significant improvement over Ref.~\cite{PI}.
The far Higgsino region is disfavored, and gaugino unification becomes
increasingly untenable as $M_2$ is taken larger than $200\gev$.  It is
also noteworthy that $\tanb>9$ for sign$(\mu)=$ sign$(M_1)$; this is
related to the fact that in this regime $D_L/D_R$ changes
significantly for low $\tanb$ when the sign of either $\mu$ or $M_1$
is changed, while changing both signs tends to cancel the effect.
Again, some correlation between $\msq$ and $\tanb$ is found, although
we will not show it here, as it is not exceptionally strong.

The allowed region is shown in Figs.~\ref{fig:Mallowedp} and
\ref{fig:Mallowedm}.  The far gaugino region is ruled out. The
Higgsino region lies outside the inner contour, as does low $\tanb$
for sign$(\mu)=$ sign$(M_1)$. (For $M_1$ negative, the symmetry
$\mu\leftrightarrow -\mu$ for large $\tanb$ only becomes obvious for
$\tanb\sim 50$.)  Unfortunately there is no overall bound on $\tanb$,
but $\tanb\sim 1$ would require that $\alpha$ lie in a small portion
of the mixed/Higgsino region.

Our results in this case study are somewhat more pessimistic
than we expect in general for this region.  We have been unable
to fully exclude the Higgsino region, because the
cross section from this case study (which depends for fixed
chargino mass on $\alpha$, $\tanb$ and $\mslep$) happens
to lie fairly close to the value of the Higgsino cross section.
Had the electron sneutrino been much lighter or heavier, the
Higgsino region would have been fully ruled out and a stronger
bound on $M_2$ would have been achieved.  Furthermore, 
we have not used the fact that two of the other neutralinos
are light enough in this case to be discovered at LEP II.
Even relatively imprecise information about the masses of
these particles could be expected to strengthen the constraints
considerably.

\section{Conclusions}
\label{sec:concl}

Inclusion of $\afbl$ and $\afbjj$ substantially enhances our estimate of
the ability of LEP II to use charginos to constrain the parameters of
the weak-scale SUSY Lagrangian.  The global bounds on the gaugino case
are quite impressive, those of the mixed case less so, and the Higgsino
case least of all, in accordance with expectations and with the results
of Ref.~\cite{PI}. Significant restrictions on $\mu/M_2$ are found in
each case study, and even $\tanb$ is constrained in the gaugino case.
No restrictions on sfermion masses are expected or found in the Higgsino
region, but light slepton masses are well-determined outside this
region, as is $M_1$.  Squark masses and, to a degree, $M_1/M_2$, are
harder to determine, but can be strongly correlated with other
quantities (as in Figs.~\ref{fig:msqminmaxp} and \ref{fig:msqminmaxm}).

It is interesting to compare the restrictions on $M_1$, $M_2$ and $\mu$
for the three cases.

\be\label{muMranges}
\begin{array}{lrccclcrccclcrcccl}
{\rm G:}&
 -1000  &<& \mu     &<&  -290 \gev  &;& 
68  &<& M_2             &<& 85\gev  &;& 
34  &<& |M_1|            &<& 44\gev \\
{\rm H:}&
 60    &<&  |\mu|     &<&  230 \gev  &;& 
75 &<& M_2             &<& 1000\gev  &;& 
60  &<& |M_1|            &<& 365\gev\\
{\rm M:}&
 55    &<&  |\mu|     &<&  175 \gev  &;& 
55 &<& M_2             &<& 560 \gev  &;& 
40  &<& |M_1|            &<& 90\gev\\
\end{array}
\ee 
where the letters refer to our gaugino, Higgsino and mixed case studies.
Note that the weakest restrictions come from the Higgsino region.
This is due in part to poorer efficiency, but there is
another more important effect at work.  The Higgsino cross section is
fixed by gauge invariance and is insensitive to other parameters.  For
a gaugino, the cross section can be very large or very small,
depending on the electron sneutrino mass $\msnu$, and therefore the 
Higgsino region often is ruled out.  By contrast, the moderately low
cross section of a Higgsino can be mimicked by a gaugino if
$\msnu$ is small\cite{Btl1,Btl2,Ahn,PI}.

We may summarize our results by reviewing the physics of the gaugino
and Higgsino regions, keeping in mind that the mixed region is in all
senses intermediate between them.  The typical characteristics of the
gaugino region --- a cross section sensitive to $\msnu$, large
chargino-neutralino mass splitting, leptonic and hadronic decays
sensitive to $\mslep$, $\msq$ and $\tanb$ --- lead to a varying number
of events, high efficiency, and strong sensitivity to all the
parameters.  Those of the Higgsino region --- a small cross section,
small chargino-neutralino mass splitting, production and decays which
are completely insensitive to $\mslep$ and $\msq$ and weakly dependent
on $\tanb$ --- lead to few events, low efficiency, and much lower
sensitivity.  The gaugino region is
characterized by a number of different sub-regions with different
properties, while there is far less variation of the physics in the
Higgsino region.  Although some parts of the gaugino region with
special characteristics mimic Higgsino physics, most do not.  Thus,
physics in the gaugino region usually rules out the Higgsino region
and some parts of the gaugino region, while physics in the Higgsino
region rules out a large fraction of the gaugino region but almost
none of the Higgsino parameter space.  This weakness of the Higgsino
case is further exacerbated by the relatively low statistics and
efficiency.  Still, even in this case it would be possible
to rule out those models which depend on gaugino mass unification
and large values of $|\mu|/M_2$. 

We note that our work is still valid (though requiring slight
reinterpretation) in the following scenarios: (1) the neutralino decays
in or out of the detector to a gravitino plus a photon or Higgs boson;
(2) other supersymmetric particles are also found at LEP II but are not
light enough to change the dominant chargino decay mode; (3) charginos
are found by the successor to LEP II at energies above $200$ GeV, but
the chargino-neutralino mass splitting is somewhat less than $M_W$.

Finally, we would like to stress two general messages from this article
and from Ref.~\cite{PI}.  Our philosophy has been oriented toward
experiment, and we have sought to keep theoretical assumptions about the
physics at very high energies out of our work.  Experimental results
from LEP II ought to be given in terms of the SUSY parameters of the
effective Lagrangian at the electroweak scale.  These weak-scale
parameters can then be related in a straightforward manner to any
particular GUT-scale or Planck-scale theoretical model, but the many
untestable theoretical assumptions which must be made in the process
should not be allowed to contaminate quoted experimental results.  Our
approach largely avoids this problem.

Also, we have found new ways of organizing SUSY parameter space, which
at first glance seems too large to control without theoretical
assumptions of the type we seek to avoid.  First, our assumptions
concern only weak-scale phenomenology, rather than, say, GUT-scale
theory, so our assumptions can be tested using the data itself.  This
approach is sufficient to reduce the number of parameters to a
manageable six (see Sec.~\ref{sec:assumptions}).  Second, we used
special properties of the chargino mass formula to project our results
onto the $(\alpha, \tanb)$ plane (see Sec.\ref{subsec:strategy}).  This
plane is a very powerful tool for characterizing the properties of
charginos, as it separates the different types of charginos into
different regions, and centers attention on the fundamental SUSY
parameters $\mu$, $M_2$ and $\tanb$.  Constraints on other parameters
and confidence contours can be usefully projected onto this plane.
Similar techniques might be employable if neutralinos are found first.
We believe this approach is practical and would be a useful tool for the
presentation of experimental results.

\acknowledgements

The authors gratefully acknowledge M.~Peskin for many useful
conversations.  J.L.F. acknowledges the support of a Short Term JSPS
Postdoctoral Fellowship and thanks the theory groups of KEK and CERN
for hospitality during the completion of this work.  J.L.F. was
supported in part by the Director, Office of Energy Research, Office of
High Energy and Nuclear Physics, Division of High Energy Physics of the
U.S.  Department of Energy under Contract DE--AC03--76SF00098 and in
part by the National Science Foundation under grant PHY--90--21139.
M.J.S. was supported in part by the Department of Energy under Contract
DE--FG05--90ER40559.

\appendix
\section{The Minimal Supersymmetric Standard Model}
\label{app:MSSM}

Our analysis is performed, as in Ref.~\cite{PI}, in the context of the MSSM
\cite{Reviews,GH}, the simplest extension of the standard model that
includes supersymmetry. In this subsection, we introduce the SUSY
parameters that we hope to constrain, and set our notation and
conventions.

The MSSM includes the usual matter superfields and two Higgs doublet
superfields

\begin{equation}\label{Hfields}
\hat{H}_1=\left( \begin{array}{c} \hat{H}_1^0 \\
                                  \hat{H}_1^- \end{array} \right)
\hspace{.2in} {\rm and} \hspace{.2in}
\hat{H}_2=\left( \begin{array}{c} \hat{H}_2^+ \\
                                  \hat{H}_2^0 \end{array} \right),
\end{equation}
where $\hat{H}_1$ and $\hat{H}_2$ give masses to the isospin
$-\frac{1}{2}$ and $+\frac{1}{2}$ fields, respectively.  These two
superfields are coupled in the superpotential through the term $ - \mu
\epsilon_{ij}\hat{H}_1^i\hat{H}_2^j$, where $\mu$ is the supersymmetric
Higgs mass parameter.  The ratio of the two Higgs scalar vacuum
expectation values is defined to be $\tanb \equiv \langle H^0_2 \rangle
/ \langle H^0_1 \rangle$. Soft SUSY-breaking terms
\cite{Girardello} for scalars and gauginos are included in the MSSM with

\begin{equation}\label{vsoft}
V_{\rm soft} = \sum_i m_i^2 |\phi_i |^2
+ \frac{1}{2} \left\{ \Big[M_1\tilde{B}\tilde{B} + \sum_{j=1}^{3}
M_2\tilde{W}^j\tilde{W}^j + \sum_{k=1}^{8} M_3\tilde{g}^k
\tilde{g}^k\Big]
+ {\rm h.c.}\right\}
+ \big[A {\rm\, terms\,} \big],
\ee
where $i$ runs over all scalar multiplets.  

The charginos and neutralinos of the MSSM are the mass eigenstates that
result from the mixing of the electroweak gauginos $\tilde{B}$ and
$\tilde{W}^j$ with the Higgsinos.  The charged mass terms that appear
are

\be
(\psi ^-)^T {\bf M}_{\chc} \psi^+ + {\rm h.c.},
\ee
where $(\psi^{\pm})^T = (-i\tilde{W}^{\pm}, \tilde{H}^{\pm})$ and

\be\label{chamass}
{\bf M}_{\chc} = \left( \begin{array}{cc}
 M_2                    &\sqrt{2} \, \mw\sinb  \\
\sqrt{2} \, \mw\cosb   &\mu                    \end{array} \right).
\ee
The chargino mass eigenstates are $\tilde{\chi}^+_i = {\bf
V}_{ij}\psi^+_j$ and $\tilde{\chi}^-_i = {\bf U}_{ij}\psi^-_j$, where
the unitary matrices {\bf U} and {\bf V} are chosen to diagonalize
${\bf M}_{\chc}$.  Neutral mass terms may be written as

\be
\frac{1}{2} (\psi ^0)^T {\bf M}_{\chn} \psi^0 + {\rm h.c.},
\ee
 where $(\psi^0)^T = (-i\tilde{B},-i\tilde{W}^3, \tilde{H}^0_1,
\tilde{H}^0_2)$ and

\be\label{neumass}
{\bf M}_{\chn} =
 \left( \begin{array}{cccc}
M_1             &0              &-\mz\cosb\sinw &\mz\sinb\sinw  \\
0               &M_2            &\mz\cosb\cosw  &-\mz\sinb\cosw \\
-\mz\cosb\sinw  &\mz\cosb\cosw  &0              &-\mu           \\
\mz\sinb\sinw   &-\mz\sinb\cosw &-\mu           &0     \end{array}
\right).
\ee
The neutralino mass eigenstates are $\chn_i = {\bf N}_{ij}\psi^0_j$,
where {\bf N} diagonalizes ${\bf M}_{\chn}$.  We take
all neutralino masses positive, rotating rows of {\bf N} by phase $i$ as
necessary. In order of increasing mass the four neutralinos are
labeled $\chn_1$, $\chn_2$, $\chn_3$, and $\chn_4$, and the two
charginos, similarly ordered, are $\chc_1$ and $\chc_2$. From the mass
matrices in Eqs.~(\ref{chamass}) and (\ref{neumass}), it can be seen
that in the limits $\tanb \rightarrow 0$ and $\tanb \rightarrow \infty$
there is an exact symmetry $\mu
\leftrightarrow -\mu$.

\section{Differential Chargino Decay Rates}
\label{app:EAdist}

We begin by considering the decay of charginos to hadrons, computing
the differential decay rate as a function of total hadron energy $E_h$ and
angle $\theta_h$ of the hadron momentum relative to the chargino spin.
In practice we compute it by integrating out the hadrons, computing
the energy $E_0$ and the angle $\theta_0$ of the neutralino in the
chargino decay, and using
  
\be
\frac{d\Gamma}{dE_h\ d\cos\theta_h}
=\frac{d\Gamma}{dE_0\ d\cos\theta_0}
\Big|_{E_0=\mchargino-E_h,\ \cos\theta_0=-\cos\theta_h} \ .
\ee

Beginning with the standard formula for the differential width in terms
of the matrix element, it is straightforward to derive
\be
\frac{d\Gamma}{dE_0\ d\cos\theta_0} = \frac{E_0}{32(2\pi)^4\mchargino} 
                 \int \ d\cos\theta\ d\phi\ |\mdecay|^2 \ .
\ee
Here $\theta$ and $\phi$ gives the direction of the quark momentum, 
measured in the quark-antiquark rest frame, relative to the direction of the
neutralino in the quark-antiquark rest frame.  The decay
matrix element $|\mdecay|$ is 
\be
\mdecay=4g^4 p_\mu\bar{p}_\nu X^{\mu\nu} \ ,
\ee
where
\be\begin{array}{ccc}
X^{\mu\nu}= &|D_L|^2 p_0^\mu (p_+ -\mchargino s)^\nu
+ |D_R|^2 (p_+ +\mchargino s)^\mu p_0^\nu \\
&+ \mLSP {\rm Re}\ (D_L D_R^*)
   \left[p_+^\mu s^\nu - s^\mu p_+^\mu -
                           \mchargino g^{\mu\nu}\right] \ .
\end{array}\ee
Here $p_+, p_0, p, \bar{p}$ are 
the momenta of the chargino, neutralino,
quark and antiquark in the decay, and $s$ is the chargino spin vector.
In this calculation we treat the $W$ propagator exactly
but treat the squark propagator as a point interaction, so
\be 
D_{L,R}(q) = \frac{O_{L,R}}{(p+\bar{p})^2-\mw^2} + \frac{Z_{L,R}(q)}{-\msq^2}
= \frac{O_{L,R}}{\mchargino^2(1-2e_0+r)^2-\mw^2} + \frac{Z_{L,R}(q)}{-\msq^2}
 \ ,
\ee
where 
\be \begin{array}{rcl}
O_L &\equiv& {\bf N}_{12} {\bf V}^*_{11}-\frac{1}{\sqrt{2}} 
    {\bf N}_{14} {\bf V}^*_{12} \\
O_R &\equiv& {\bf N}^*_{12} {\bf U}_{11}+\frac{1}{\sqrt{2}} 
    {\bf N}^*_{13} {\bf U}_{12}
\end{array}
\ee
and
\be\begin{array}{rcl}
Z_L(q) &\equiv& 
   {\bf V}^*_{11}(\frac{1}{6} \tanw {\bf N}_{11}+\half {\bf N}_{12})\\
Z_R(q) &\equiv& - 
{\bf U}_{11}(\frac{1}{6} \tanw {\bf N}^*_{11}-\half {\bf N}^*_{12}) \ ,
\end{array}\ee
and where we define

\be
e_0=\frac{E_0}{\mchargino} \qquad ; \qquad r =
\left(\frac{\mLSP}{\mchargino}\right)^2 \ .
\ee

The integration over $\theta$ and
$\phi$ is trivial since $X^{\mu\nu}$ is independent of those angles,
and thus

\be
\frac{d\Gamma}{dE_0\ d\cos\theta_0} = 
\frac{g^4 \mchargino^4}{48(2\pi)^3}
\frac{\sqrt{e_0^2-r}}{\left[\mw^2-\mchargino^2(1-2e_0+r)\right]^2}
\left[f_0(e_0,r) + f_1(e_0,r) \cos\theta_0 \right] \ ,
\ee
where

\be\begin{array}{rcl}
f_0 &=& (|D_R|^2+|D_L|^2)\left[3e_0(1+r)-2r-4e_0^2\right]
       -6 \sqrt{r}{\rm Re}\ (D_L D_R^*)(1-2e_0+r)
\\ \\
f_1 &=& (|D_R|^2-|D_L|^2)\sqrt{e_0^2-r}\left[4e_0-1-3r\right] \ .
\end{array}\ee

Next we turn to the differential decay rate in chargino
decay via leptons as a function of the energy $\bar{E}$ and
angle $\bar\theta$ of the charged antilepton.  Again we treat
the $W$ propagator exactly and the squark propagator as a point
interaction.  However, in this case the formulas
are more complicated because the $W$ propagator depends on the angle
between the charged antilepton and the neutrino.
Again, simple manipulations lead to
\be
\frac{d\Gamma}{d\bar{E}\ d\cos\bar{\theta}} 
= \frac{\bar{E}}{32(2\pi)^4 \mchargino} F(\bar{e},r)
       \int  d\cos\theta\ d\phi\ |\mdecay|^2 \ ,
\ee
using notation as in the previous case along with
\be
\bar{e} = \frac{\bar{E}}{\mchargino} \qquad ; \qquad
 F(\bar{e},r) = \frac{1-2\bar{e}-r}{1-2\bar{e}} \ .
\ee
Here $\theta$ and $\phi$ gives the direction of the neutrino momentum, 
measured in the neutrino-neutralino rest frame, 
relative to the direction of the
charged antilepton in the neutrino-neutralino rest frame.  The decay
matrix element $|\mdecay|$ is 
\be
\mdecay=4g^4\left[Y_L^{\mu\nu}I_{L\mu\nu} + Y_R^{\mu\nu}I_{R\mu\nu} 
\right] + V^\mu J_\mu \ ,
\ee
with
\be\begin{array}{rcl}
Y_L^{\mu\nu}&= & \bar{p}\cdot(p_+ -\mchargino s)g^{\mu\nu} \\
Y_R^{\mu\nu}&= & (p_+ +\mchargino s)^\mu \bar{p}^\nu \\ 
V^\mu &= & \mLSP  
   \left[p_+^\mu \bar{p}\cdot s - s^\mu \bar{p}\cdot p_+ -
                           \mchargino \bar{p}^\mu\right] \\ \\
I_{L\mu\nu}&= & \int d\cos\theta\ d\phi\  |D_L|^2 p_\mu p_{0\nu} \\
I_{R\mu\nu}&= & \int d\cos\theta\ d\phi\  |D_R|^2 p_\mu p_{0\nu} \\
J_\mu &= & \int d\cos\theta\ d\phi\  {\rm Re}\ (D_L D_R^*) p_\mu \ .
\end{array}\ee
The expressions for $D_L$ and $D_R$ are
\be 
D_{L,R}(l) = \frac{O_{L,R}}{(p+\bar{p})^2-\mw^2} + 
\frac{Z_{L,R}(l)}{-\mslep^2}
= \frac{O_{L,R}}{-\mw^2\left[1-\rho(1-\cos\theta)\right]} + 
\frac{Z_{L,R}(l)}{-\mslep^2} \ ,
\ee
where $O_{L,R}$ are as above,
\be\
\rho \equiv  \frac{\mchargino^2}{\mw^2}\bar{e}F(\bar{e},r) \ ,
\ee
and
\be\begin{array}{rcl}
Z_L(l) &\equiv& 
   {\bf V}^*_{11}(-\frac{1}{2} \tanw {\bf N}_{11}+\half {\bf N}_{12})\\
Z_R(l) &\equiv& - 
{\bf U}_{11}(-\frac{1}{2} \tanw {\bf N}^*_{11}-\half {\bf N}^*_{12}) \ .
\end{array}\ee

This leads to the result
\be
\frac{d\Gamma}{d\bar{E}\ d\cos\bar{\theta}} 
= \frac{g^4 \mchargino^4\bar{e}^2}{16(2\pi)^3 \mw^4} 
\left[F(\bar{e},r)\right]^2 
\left[h_0(\bar{e},r)(1+\cos\bar{\theta})
+h_1(\bar{e},r)(1-\cos\bar{\theta})
+h_2(\bar{e},r)\right] \ ,
\ee
where
\be\begin{array}{rcl}
h_0 &=& (1-2\bar{e})H_{LL}^{(0)} - \sqrt{r}H_{RL}^{(1)}
\\ \\
h_1 &=& \left[(1-2\bar{e})H_{RR}^{(0)}-(1-\bar{e})H_{RR}^{(1)} \right]
  -\half F(\bar{e},r)
  \left[(1-2\bar{e})H_{RR}^{(1)}-(1-\bar{e})H_{RR}^{(2)}\right]
\\ \\
h_2 &=& H_{RR}^{(1)} -\half
F(\bar{e},r)H_{RR}^{(2)}
 \ . \end{array}\ee
The functions $H_{AB}^{(n)}$, for $A,B = R,L$ and $n=0,1,2$, are
\be
H_{AB}^{(n)} = {\rm Re}\left\{O_A O_B^* P_2^{(n)}(\rho) + 
            \frac{\mw^2}{\mslep^2}
               \left[O_A Z_B^*(l)+ Z_A(l) O_B^*\right] P_1^{(n)}(\rho)
       +  \frac{\mw^4}{\mslep^4}
                         Z_A(l) Z_B^*(l) P_0^{(n)}(\rho)\right\} \ ,
\ee
where 
\be
P_k^{(n)} 
= \int_{-1}^1 \frac{\ds d\cos\theta\ (1-\cos\theta)^n} 
{\ds [1-\rho (1-\cos\theta)]^k}
\ .
\ee

In the limit $r\rarr 1$, the angular dependence
of the $W$-boson propagator can be ignored, 
so this result simplifies to 
\be\label{afbldGam}
\frac{d\Gamma}{d\bar{E}\ d\cos\bar{\theta}} 
= \frac{g^4 \mchargino^4\bar{e}^2}{8(2\pi)^3 \mw^4} 
\left[F(\bar{e},r)\right]^2 \left[f_0(\bar{e},r) + f_1(\bar{e},r) 
\cos\bar{\theta} \right] \ ,
\ee
where
\be\begin{array}{rcl}
f_0 &=& |D_L|^2(1-2\bar{e}) - {\rm Re}(D_R D_L^*)\sqrt{r}
 + |D_R^2|\left\{1-\bar{e}-\frac{3-2\bar{e}}{6}F(\bar{e},r)\right\}
\\ \\
f_1 &=& |D_L|^2(1-2\bar{e}) - {\rm Re}(D_R D_L^*)\sqrt{r}
 + |D_R^2|\left\{\bar{e}-\frac{1+2\bar{e}}{6}F(\bar{e},r)\right\} \ .
\end{array}\ee

\input psfig

\noindent
\begin{figure}\hspace*{1in}
\psfig{file=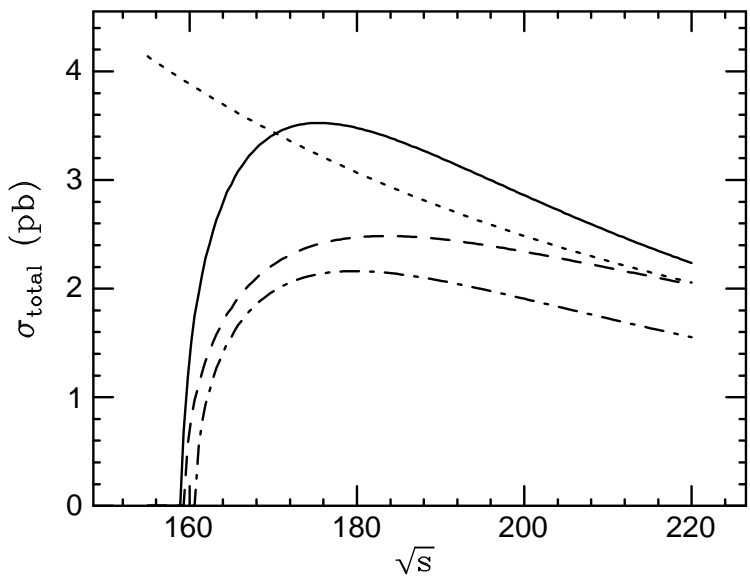,width=0.6\textwidth} \vspace{.1in}
\caption{\label{fig:csec}
  The chargino cross section as a function of center-of-mass energy for
  the three choices of parameters of our case studies: the gaugino case
  [solid line, see Eq.~(\protect\ref{gparam})], the Higgsino case
  [dashed line, see Eq.~(\protect\ref{hparam})], and the mixed case
  [dot-dashed line, see Eq.~(\protect\ref{mparam})].  A unit of R is
  also shown (dotted line). }
\end{figure}

\noindent 
\begin{figure}\hspace*{.7in}
\psfig{file=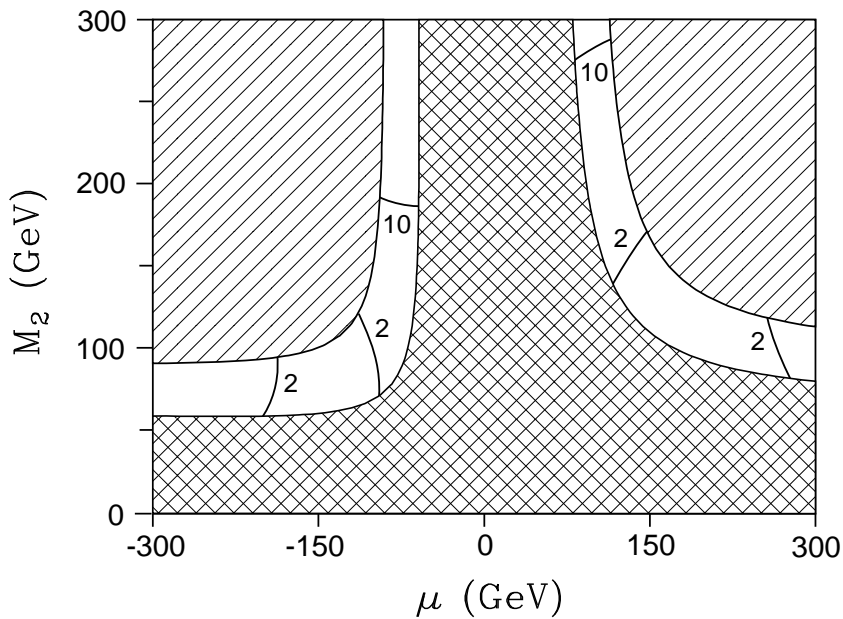,width=0.7\textwidth} \vspace{.1in}
\caption{\label{fig:sigmalr}
    Contours of constant $\sigma_R/\sigma_L$ in percent in the $\mum$
    plane for $\tanb=4$ and a sneutrino mass $\msnu=150 \gev$, for which
    the ratio is nearly maximal.  The ratio never rises above 15\% in
    the allowed bands, and is approximately 2\% in much of the gaugino
    region. The cross-hatched region is excluded by the current chargino
    mass bound $\mchargino > 65 \gev$ \protect\cite{LEP1.3}, while in
    the hatched regions, $\mchargino > 95 \gev$, so charginos are
    kinematically inaccessible at LEP II.}
\end{figure}

\noindent 
\begin{figure}\hspace*{.7in}
\psfig{file=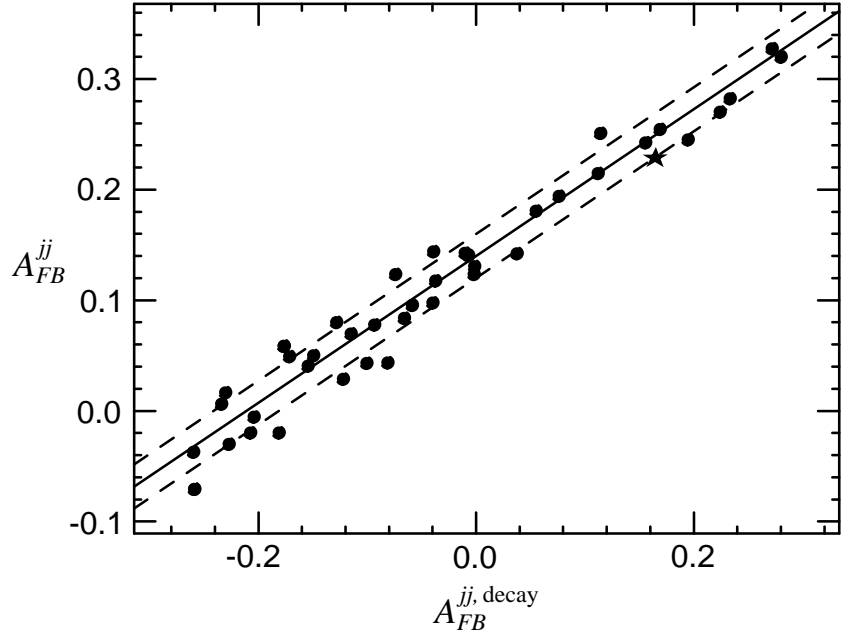,width=0.7\textwidth} \vspace{.1in}
\caption{\label{fig:afbcor}
  Plot of observed $\afbjj$ vs.~decay amplitude asymmetry $\afbjjd$ for
  numerous points with the same $\mchargino$, $\mLSP$, $\stot$ and $B_l$
  as the gaugino case study, which is indicated by a star.  A linear
  fit is shown by the solid line with the standard deviation for the fit
  indicated by the dashed lines.}
\end{figure}

\noindent 
\begin{figure}\hspace*{.7in}
\psfig{file=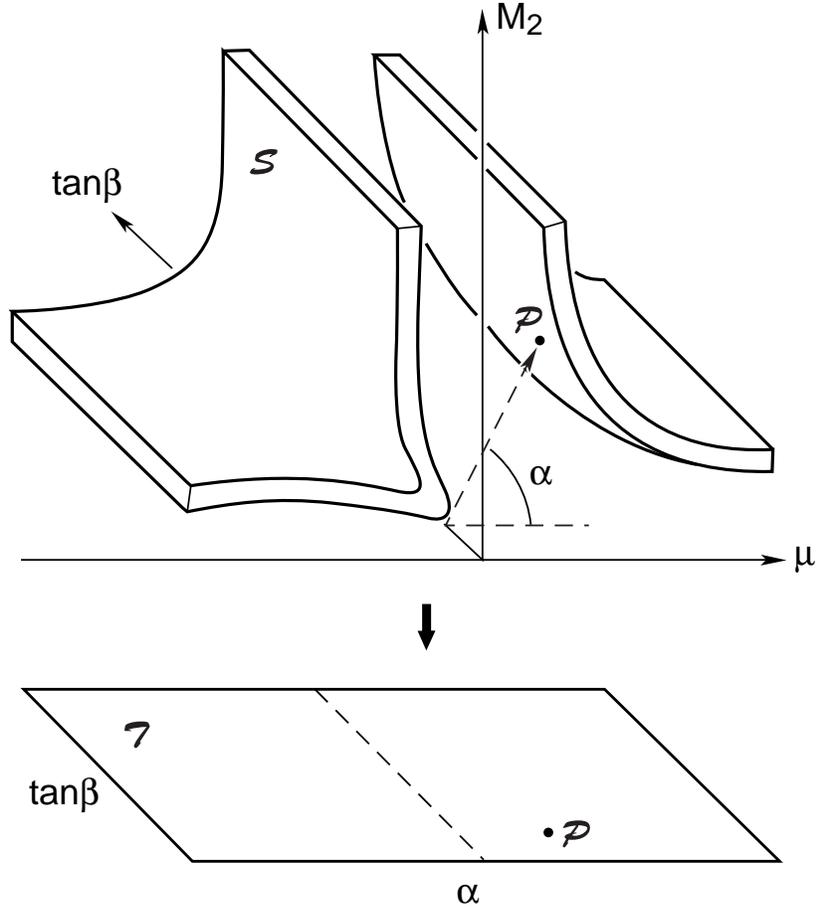,width=0.7\textwidth} \vspace{.1in}
\caption{\label{fig:sheets}
  The $\mchargino$ measurement restricts the $(\mu, M_2, \tanb)$ space
  to two thin sheets $\cal S$, which are then flattened into the plane
  $\cal T$ with the transformation $(\mu, M_2, \tanb) \rightarrow
  (\alpha, \tanb)$, where $\alpha = \arctan (M_2/\mu)$. This
  transformation is illustrated schematically here.  For large $\tanb$,
  observables are symmetric under $\mu\leftrightarrow -\mu$, that is,
  under $\alpha\leftrightarrow 180^\circ-\alpha$.}
\end{figure}

\noindent 
\begin{figure}\hspace*{.7in}
\psfig{file=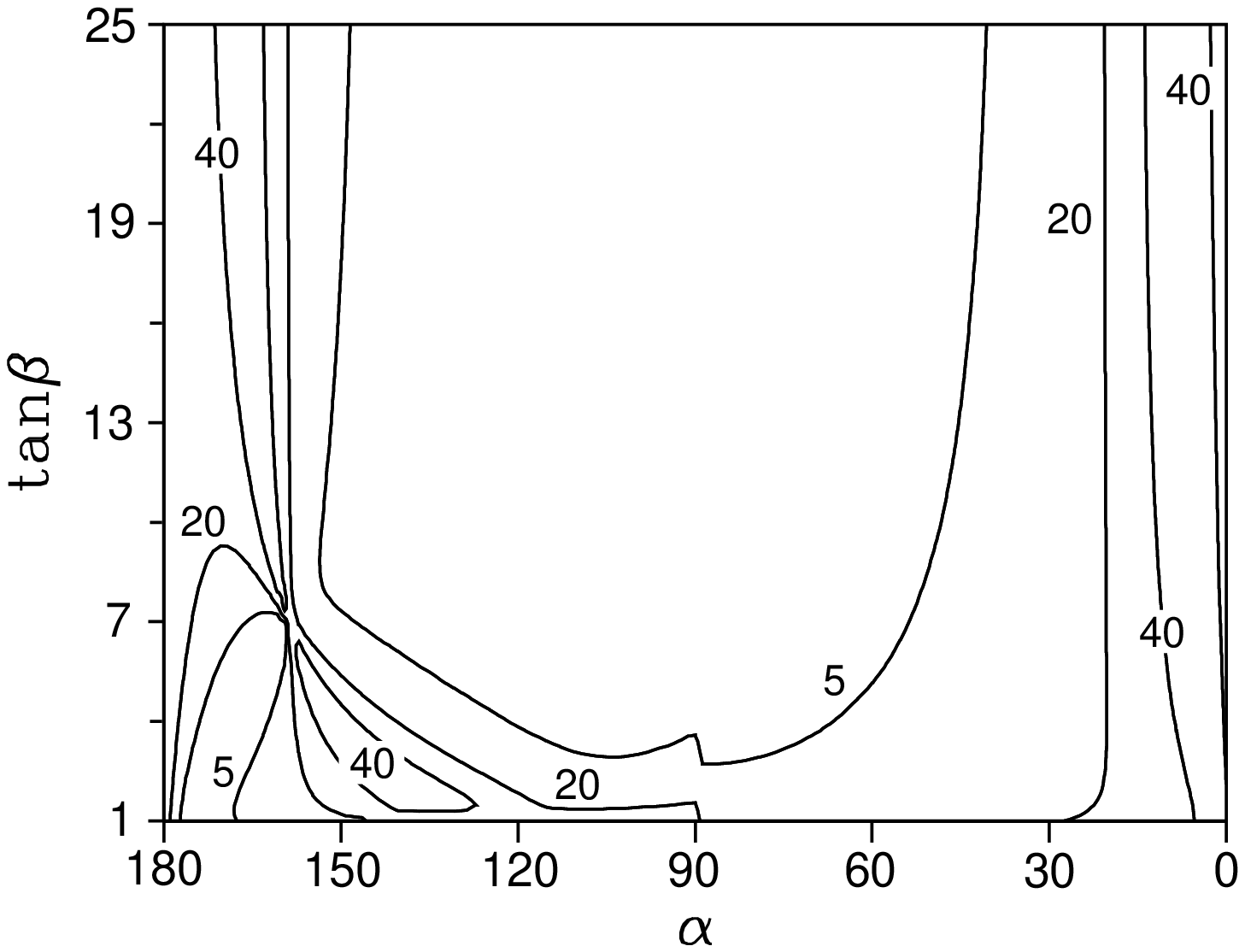,width=0.7\textwidth} \vspace{.1in}
\caption{\label{fig:afblG} Contours of $\afbld$ in percent for 
  $M_1>0$ and $\mchargino=80\gev$, $\mLSP = 40\gev$,
  $\mslep=200\gev$, plotted in the $(\alpha, \tanb )$ plane, as
  defined in Sec.~\protect\ref{subsec:strategy}.}
\end{figure}

\newpage
\noindent 
\begin{figure}\hspace*{.7in}
\psfig{file=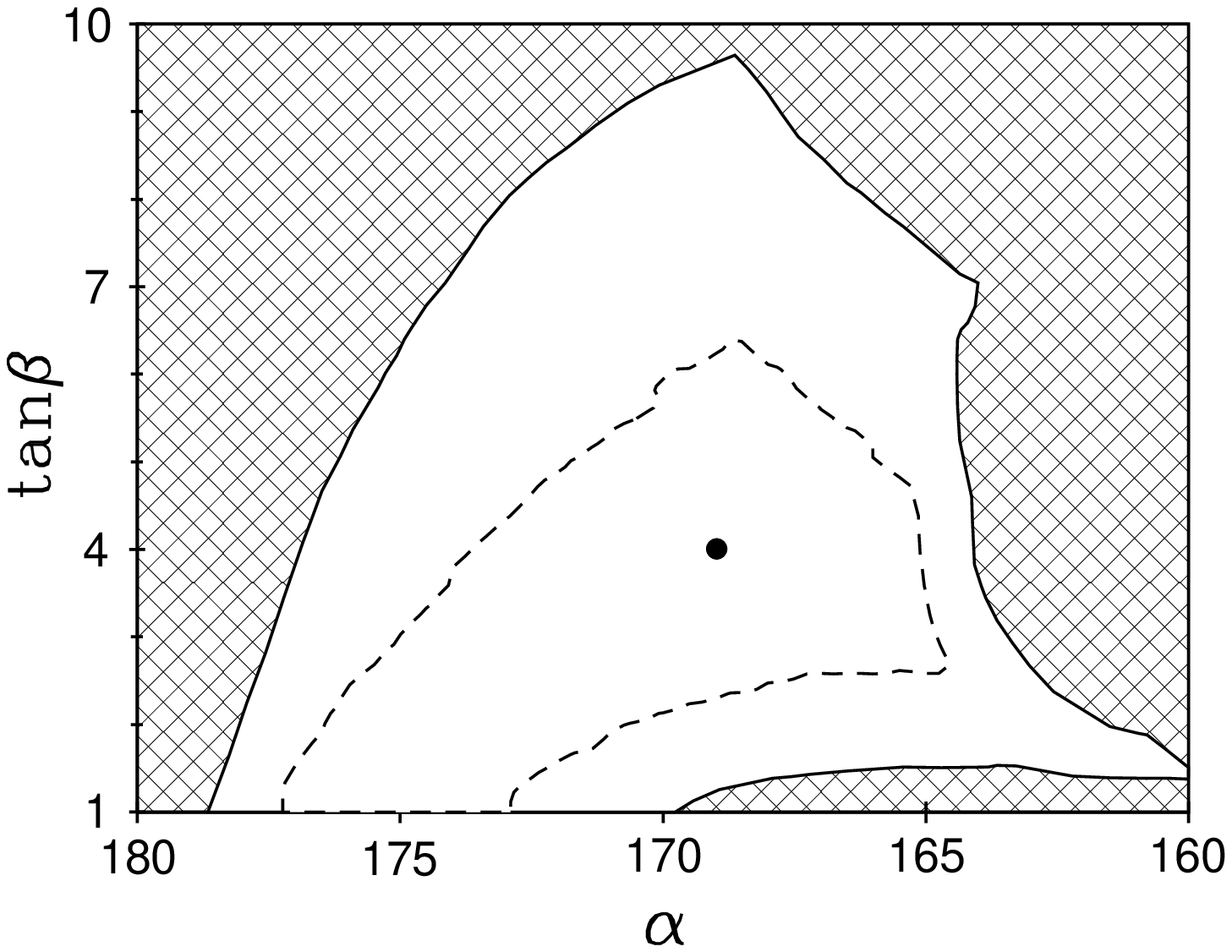,width=0.7\textwidth} \vspace{.1in}
\caption{\label{fig:Gallowedp} Plot of the allowed region 
  (defined in Sec.~\protect\ref{subsec:strategy}) for the gaugino case
  study and $M_1>0$.  The dashed (solid) contour is the projection of
  the region of parameter space in which all observables are within
  one (two) standard deviation(s) of their central values; see
  Sec.~\protect\ref{subsec:strategy} for discussion.  The dot
  indicates the value of $(\alpha, \tanb )$ for the case study.}
\end{figure}

\noindent 
\begin{figure}\hspace*{.7in}
\psfig{file=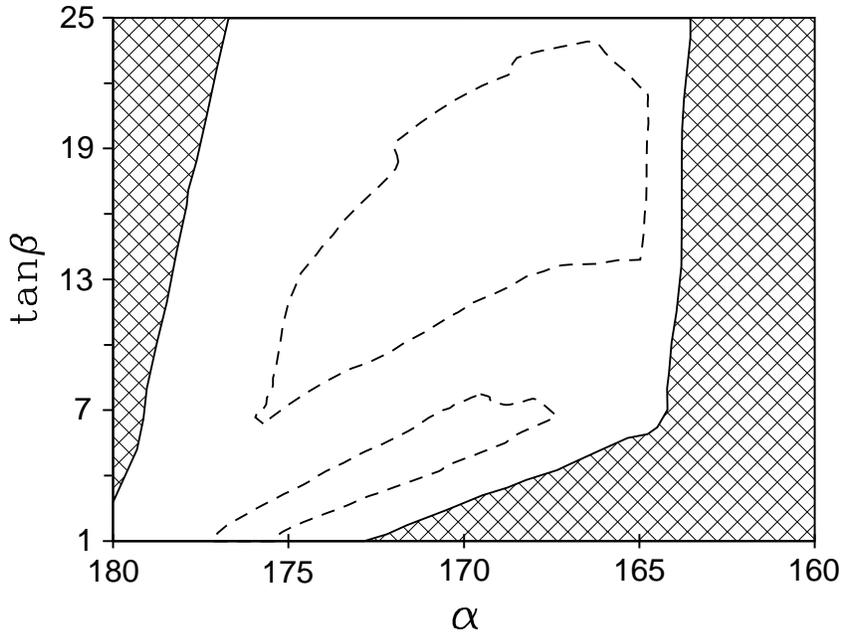,width=0.7\textwidth} \vspace{.1in}
\caption{\label{fig:Gallowedm} Same as Fig.~\protect\ref{fig:Gallowedp} 
   but for $M_1<0$.  The outer region (solid contour) extends to
  $\tanb=38$.}
\end{figure}

\noindent 
\begin{figure}\hspace*{.7in}
\psfig{file=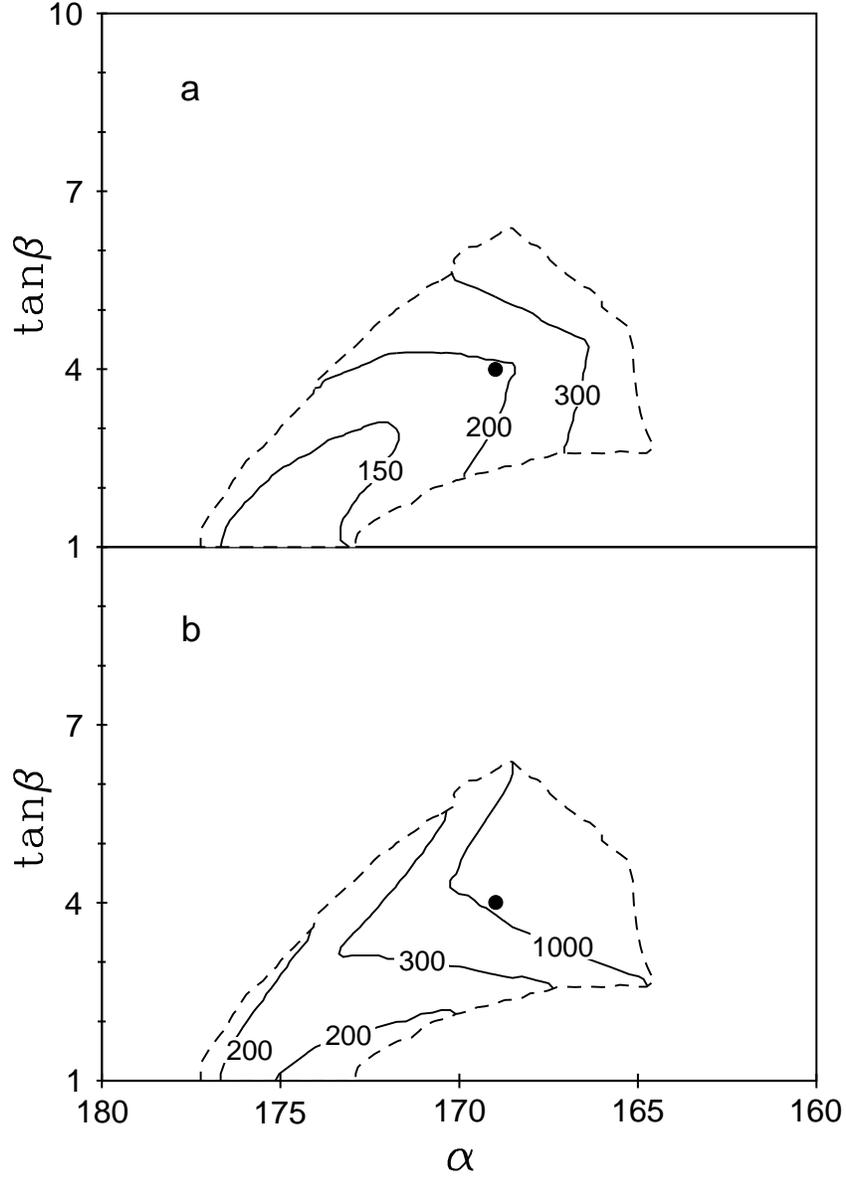,width=0.7\textwidth} \vspace{.1in}
\caption{\label{fig:msqminmaxp}
  Contours of the (a) minimum and (b) maximum values of $\msq$ in the
  $(\alpha, \tanb )$ plane, as defined in
  Sec.~\protect\ref{subsec:strategy}, for the gaugino case study and
  $M_1>0$, shown only inside the inner allowed region.  This
  illustrates the correlation between $\msq$, $\alpha$ and $\tanb$.
  The dot indicates the value of $(\alpha, \tanb )$ for the case
  study.}
\end{figure}

\noindent 
\begin{figure}\hspace*{.7in}
\psfig{file=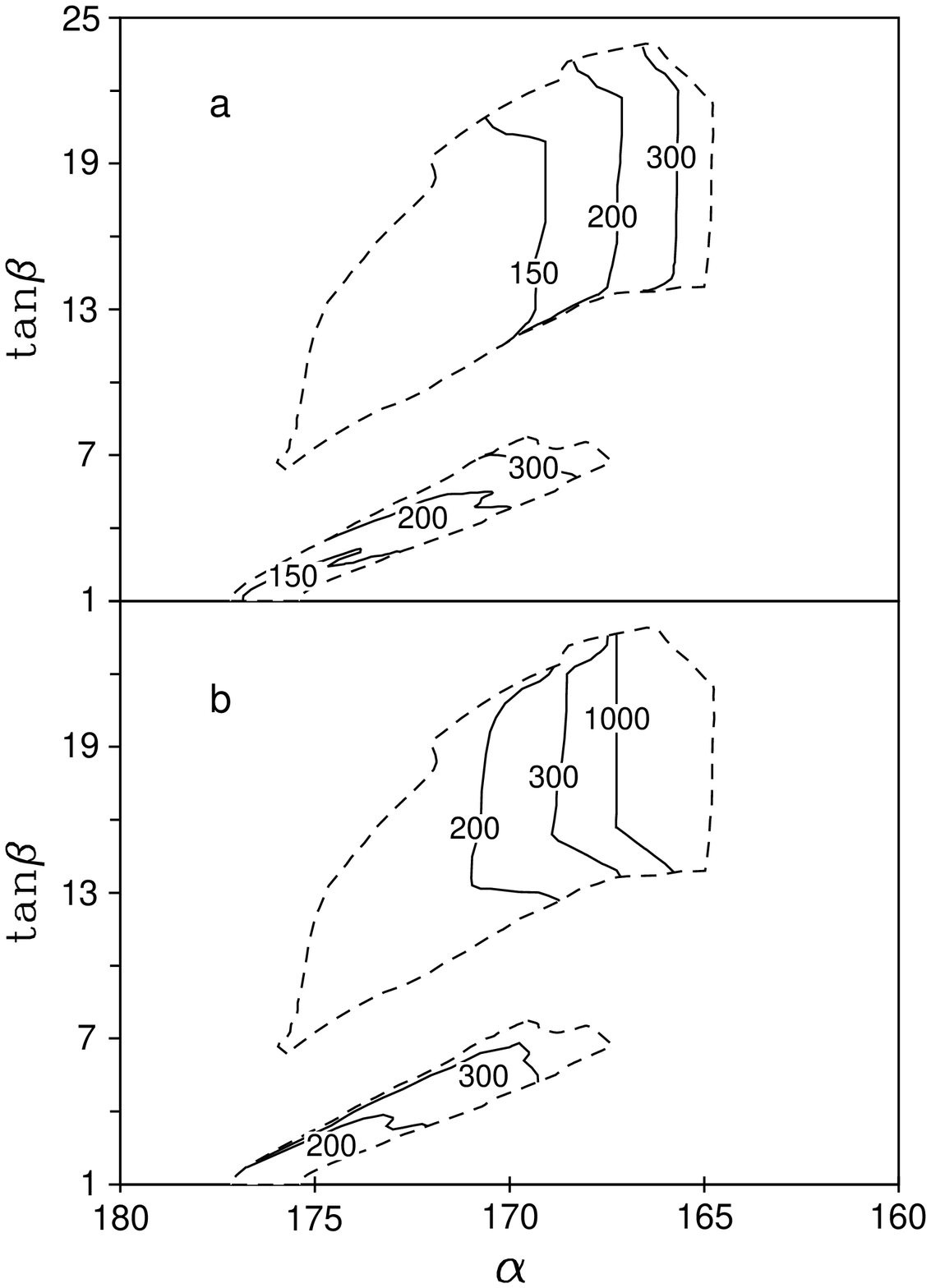,width=0.7\textwidth} \vspace{.1in}
\caption{\label{fig:msqminmaxm} Same as
  Fig.~\protect\ref{fig:msqminmaxp} but for $M_1<0$.}
\end{figure}

\noindent 
\begin{figure}\hspace*{.7in}
\psfig{file=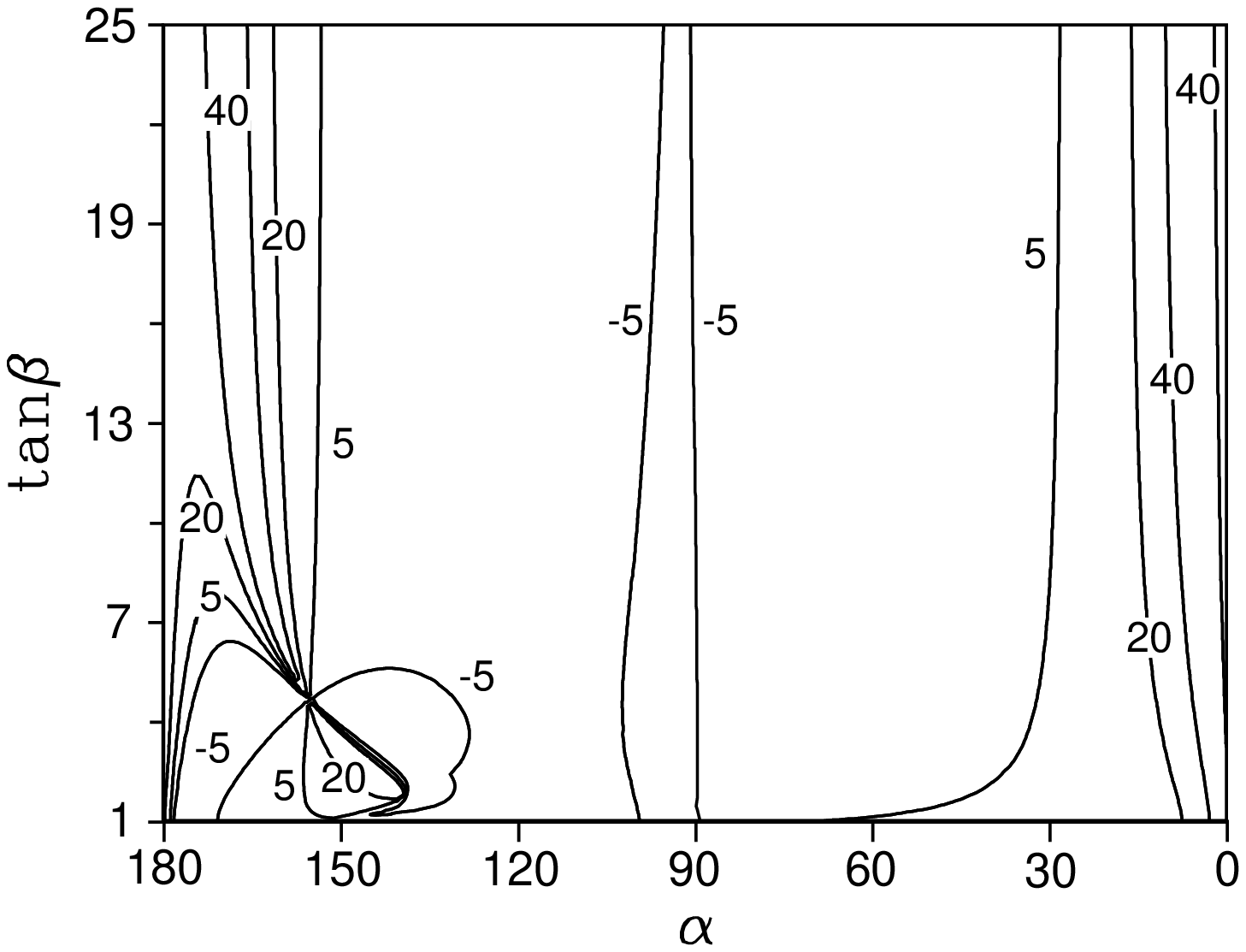,width=0.7\textwidth} \vspace{.1in}
\caption{\label{fig:afblH}Contours of $\afbld$ in percent for 
  $M_1>0$ and $\mchargino=80\gev$, $\mLSP = 62\gev$,
  $\mslep=175\gev$, plotted in the $(\alpha, \tanb )$ plane, as
  defined in Sec.~\protect\ref{subsec:strategy}.}
\end{figure}

\noindent 
\begin{figure}\hspace*{.7in}
\psfig{file=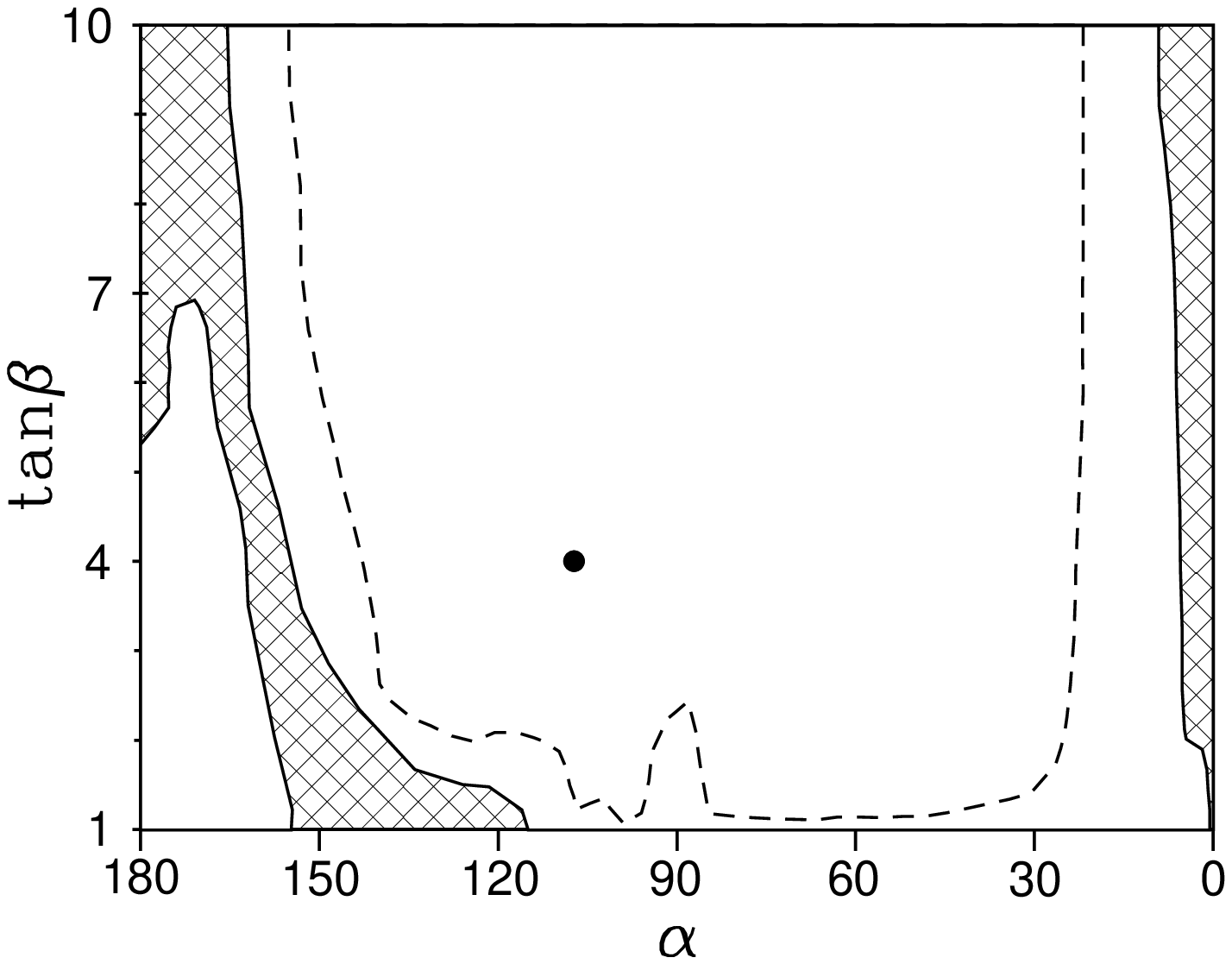,width=0.7\textwidth} \vspace{.1in}
\caption{\label{fig:Hallowedp}Plot of the allowed region 
  (defined in Sec.~\protect\ref{subsec:strategy}) for the Higgsino
  case study and $M_1>0$.  The dashed (solid) contour is the
  projection of the region of parameter space in which all observables
  are within one (two) standard deviation(s) of their central values;
  see Sec.~\protect\ref{subsec:strategy} for discussion. The dot
  indicates the value of $(\alpha, \tanb )$ for the case study.}
\end{figure}

\noindent 
\begin{figure}\hspace*{.7in}
\psfig{file=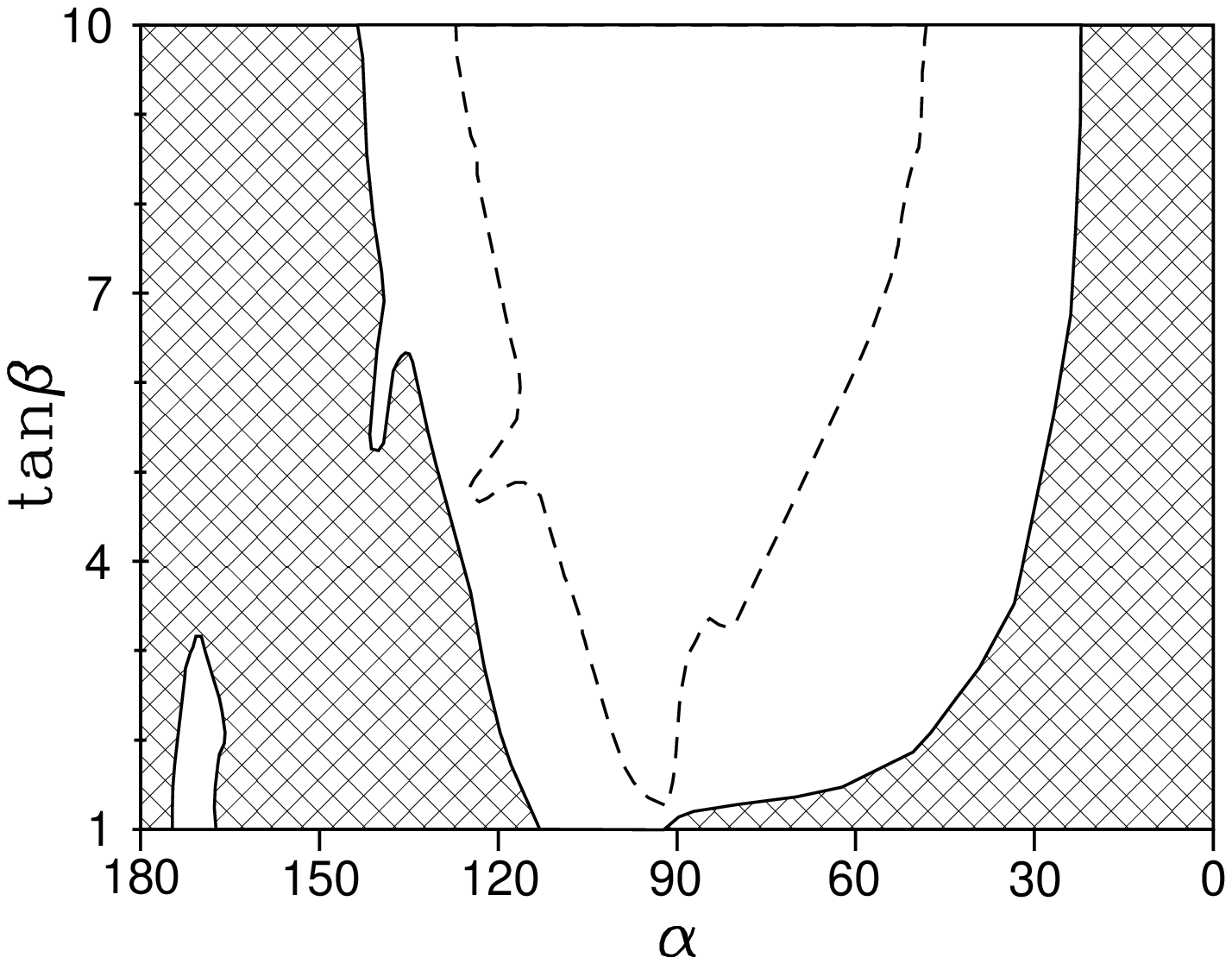,width=0.7\textwidth} \vspace{.1in}
\caption{\label{fig:Hallowedm} Same as Fig.~\protect\ref{fig:Hallowedp} 
  but for $M_1<0$.}
\end{figure}

\noindent 
\begin{figure}\hspace*{.7in}
\psfig{file=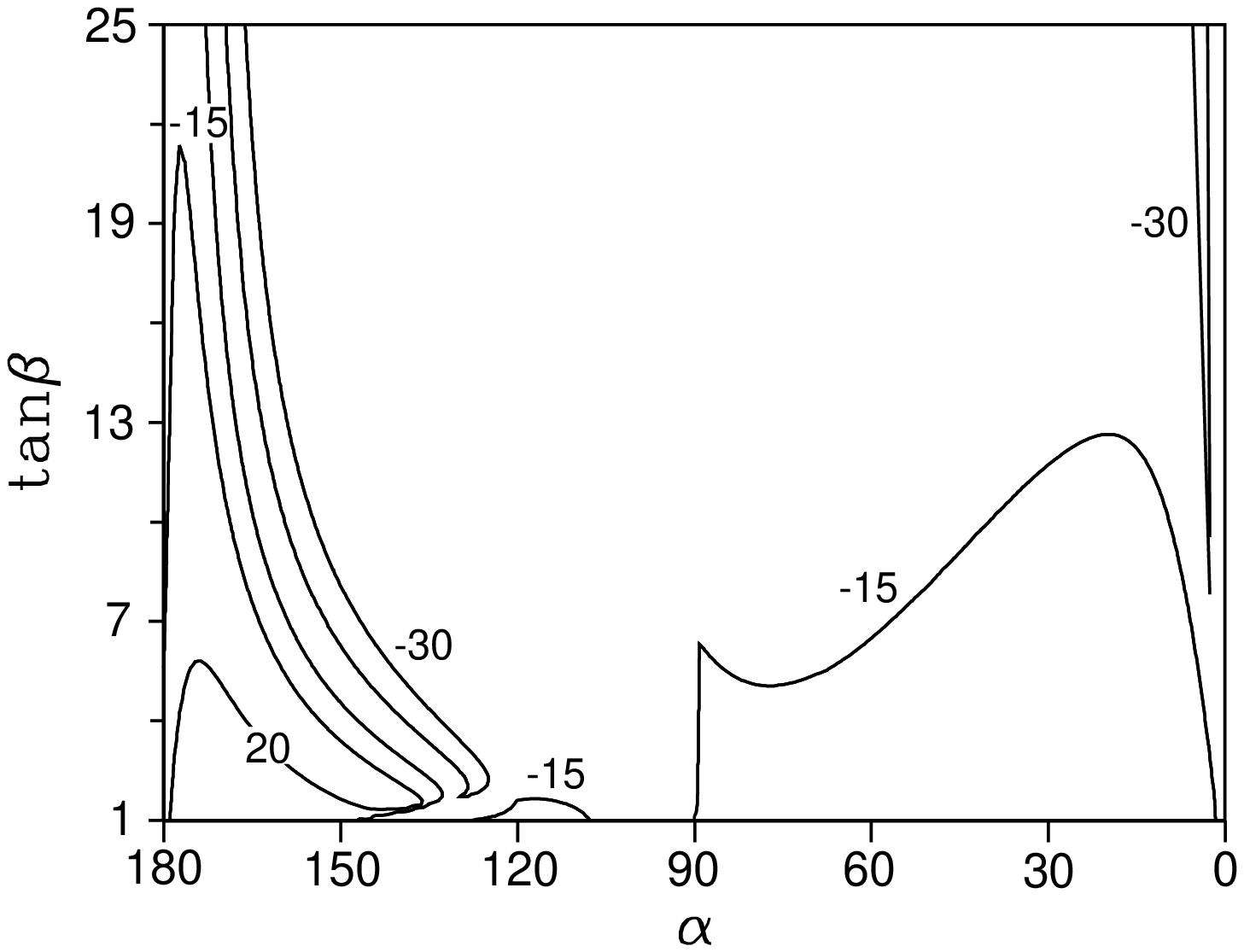,width=0.7\textwidth} \vspace{.1in}
\caption{\label{fig:afbjjM}Contours of $\afbjjd$ in percent for 
  $M_1>0$ and $\mchargino=80\gev$, $\mLSP = 53\gev$, $\msq=300\gev$,
  plotted in the $(\alpha, \tanb )$ plane, as defined in
  Sec.~\protect\ref{subsec:strategy}.  Jumps in the contours across
  $\alpha=90^{\circ}$ are due to the discontinuity in the underlying
  parameters across this line; see Fig.~\protect\ref{fig:sheets}.}
\end{figure}

\noindent 
\begin{figure}\hspace*{.7in}
\psfig{file=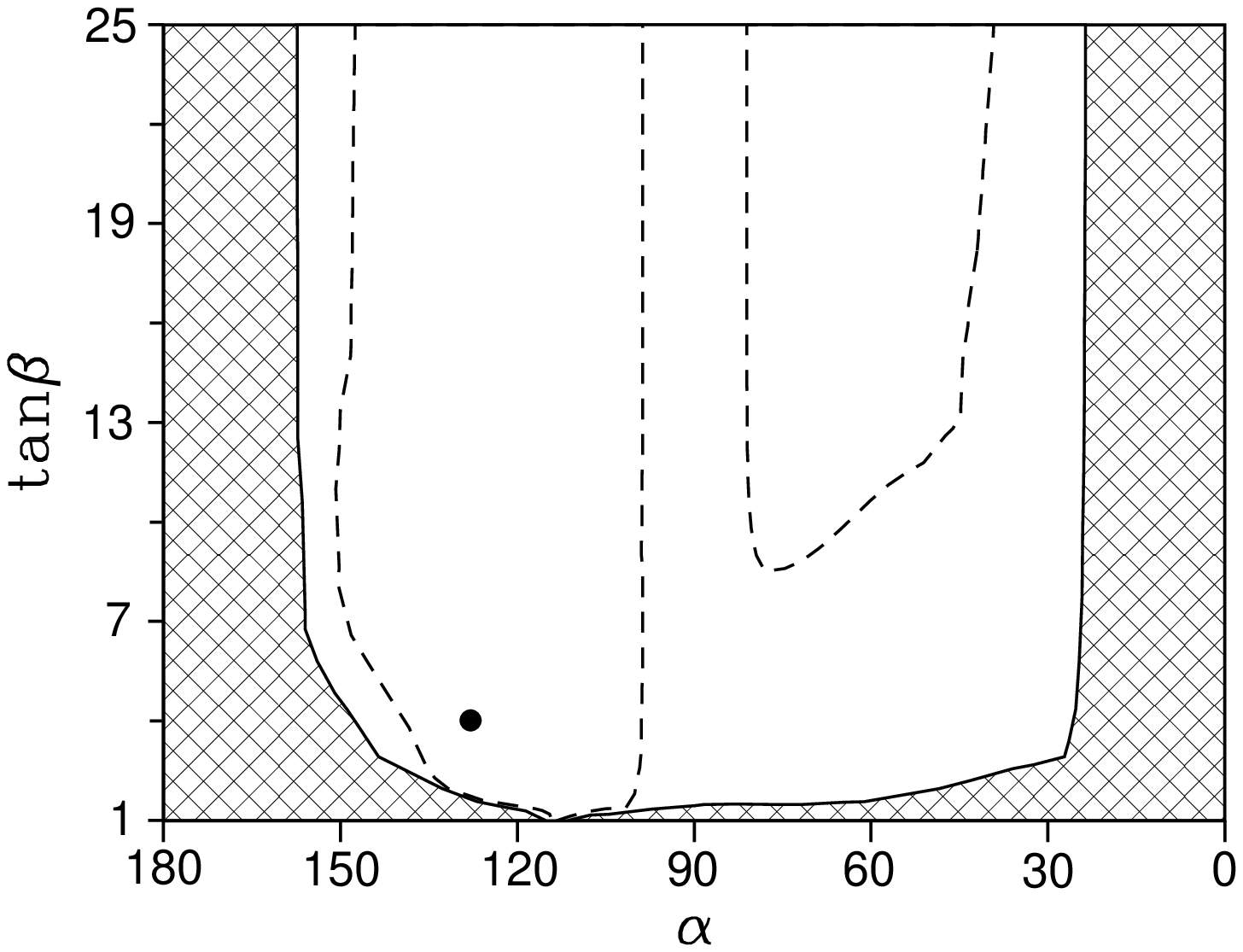,width=0.7\textwidth} \vspace{.1in}
\caption{\label{fig:Mallowedp}Plot of the allowed region 
  (defined in Sec.~\protect\ref{subsec:strategy}) for the mixed case
  study and $M_1>0$.  The dashed (solid) contour is the projection of
  the region of parameter space in which all observables are within
  one (two) standard deviation(s) of their central values; see
  Sec.~\protect\ref{subsec:strategy} for discussion. The dot indicates
  the value of $(\alpha, \tanb )$ for the case study.}
\end{figure}

\noindent  
\begin{figure}\hspace*{.7in} 
\psfig{file=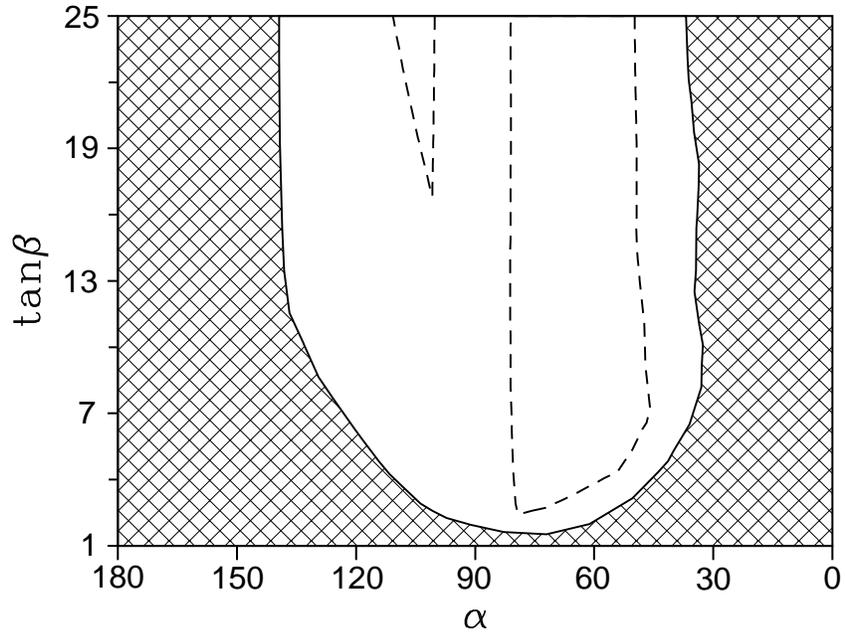,width=0.7\textwidth} \vspace{.1in} 
\caption{\label{fig:Mallowedm} Same as Fig.~\protect\ref{fig:Mallowedp} 
  but for $M_1<0$.  The approximate symmetry $\alpha \leftrightarrow
  180^{\circ}-\alpha\ $ ($\mu \leftrightarrow -\mu$) for large $\tanb$
  only becomes evident near $\tanb=50$.}
\end{figure}

\end{document}